\documentclass[review]{elsarticle}
\usepackage{tabularx}
\usepackage[utf8]{inputenc}
\usepackage{lineno,hyperref}
\usepackage[top=1.65in, bottom=1.65in, left=1.35in, right=1.35in]{geometry}
\usepackage{setspace}
\usepackage{amsmath, amssymb}

\usepackage{caption} 
\usepackage{graphicx}
\usepackage{algorithm}
\usepackage[noend]{algpseudocode}
\usepackage{subcaption}
\usepackage{geometry}
\usepackage{listings}
\usepackage{xcolor}

\definecolor{codebg}{rgb}{0.95,0.95,0.95}
\definecolor{keywordcolor}{rgb}{0.0,0.0,0.6}
\definecolor{commentcolor}{rgb}{0.3,0.3,0.3}
\definecolor{stringcolor}{rgb}{0.5,0.0,0.0}

\biboptions{numbers}

\makeatletter
\def\ps@pprintTitle{%
 \let\@oddhead\@empty
 \let\@evenhead\@empty
 \def\@oddfoot{}%
 \let\@evenfoot\@oddfoot}
\makeatother
\begin{document}

\begin{frontmatter}

\title{Augmenting the Interpretability of GraphCodeBERT for Code Similarity Tasks}
\author{Jorge Martinez-Gil}
\address{Software Competence Center Hagenberg GmbH \\ Softwarepark 32a, 4232 Hagenberg, Austria \\ \url{jorge.martinez-gil@scch.at}}

\begin{abstract}
Assessing the degree of similarity of code fragments is crucial for ensuring software quality, but it remains challenging due to the need to capture the deeper semantic aspects of code. Traditional syntactic methods often fail to identify these connections. Recent advancements have addressed this challenge, though they frequently sacrifice interpretability. To improve this, we present an approach aiming to improve the transparency of the similarity assessment by using GraphCodeBERT, which enables the identification of semantic relationships between code fragments. This approach identifies similar code fragments and clarifies the reasons behind that identification, helping developers better understand and trust the results. The source code for our implementation is available at \url{https://www.github.com/jorge-martinez-gil/graphcodebert-interpretability}.
\end{abstract}


\end{frontmatter}

\section{Introduction}
The growing complexity of software systems requires appropriate methods for analyzing code, particularly those that can recognize and compare code on a deeper level beyond syntax \cite{wei2017supervised}. Traditional strategies tend to focus on surface-level similarity, which can result in missing important aspects, especially in cases where code is written using different styles. This means that, without a deeper understanding, these methods may yield incomplete or misleading results, impacting the effectiveness of code maintenance and optimization efforts within increasingly large and complex projects.

To face these problems, transformer models like GraphCodeBERT \cite{guo2020graphcodebert} have undergone pre-training on extensive datasets of source code. These models provide a promising solution for a more effective exploration of the semantic structure of code. They use a transformer architecture to analyze relationships and meaning. The idea behind learning from diverse examples across various programming languages allows GraphCodeBERT to identify the intent and functionality behind different code fragments, even when those fragments appear different on the surface.

However, the complexity of these models poses a challenge, as their vast architectures and encoded multi-dimensional information make them difficult for humans to interpret \cite{abs-2407-02646}. It is, therefore, crucial to explore methods that improve our understanding of these models \cite{CheferGW21}. This work faces the interpretability challenge by introducing a GraphCodeBERT-driven framework that visualizes the model's inner workings while determining the similarity between code fragments. The key contributions of this research include:

\begin{itemize}
    \item We introduce a new approach using GraphCodeBERT to visualize the semantic similarity between two code fragments. This approach breaks down the code into tokens and embeds them into high-dimensional vectors to assess the significance of each token relative to the others.
    
    \item The approach produces visual representations that illustrate how closely the tokens in two code fragments are related in terms of meaning. Our strategy proves particularly useful in assessing code quality, detecting potential plagiarism, or identifying opportunities for refactoring.
    
    \item Furthermore, this method has potential applications across various aspects of software engineering. For example, it can suggest code improvements or improve automated code generation systems. We aim to demonstrate how GraphCodeBERT's capabilities can contribute to developing more efficient and maintainable software systems.
\end{itemize}

The rest of our work guides the reader through the research: Section 2 examines existing methods and places this work within the context of related research. Section 3 outlines the approach, from tokenization to creating similarity matrices. Section 4 presents a use case for applying the proposed strategy to classical sorting algorithms. Section 5 discusses the strengths, limitations, and potential future lines of research. The conclusion summarizes the key contributions and emphasizes the importance of the findings.

\section{State-of-the-Art}
Researchers have made significant progress in code representation, especially with the introduction of pre-trained models for programming languages. These models, primarily based on transformer architectures, achieve strong performance in tasks such as code completion \cite{bruch2009learning} and similarity detection \cite{key-martinez-mlwa}. This section reviews the current approaches relevant to our work.

\subsection{Pre-trained Models for Code Understanding}
Pre-trained models have changed the field of automatic text processing. The idea is to use large-scale unsupervised training on vast amounts of textual data \cite{key-Bert}. Building on these advances, researchers applied similar techniques to programming languages \cite{key-codebert}.

CodeBERT \cite{key-codebert}, the seminal model in this area, uses a bi-directional transformer pre-trained on large datasets of source code. It is designed to learn both syntactic and semantic aspects of code and has been fine-tuned for tasks such as code repair \cite{Mashhadi2021} and clone detection \cite{ain2019systematic}.

An extension of CodeBERT, GraphCodeBERT \cite{guo2020graphcodebert}, includes additional structural information. Incorporating data flow graphs, representing the relationships between variables, allows for capturing deeper semantic details, improving its performance for tasks requiring a detailed understanding of code behavior. Models of this kind have many applications, for example, identification of code vulnerabilities \cite{WangXTJSJ24}.

\subsection{Tokenization and Embedding Techniques}
Tokenization is a critical step in preparing code for machine learning models. Traditional tokenization methods used in text processing often struggle with programming languages due to their unique syntax and use of various symbols and operators. More recent models use specialized tokenization techniques better suited for code, allowing them to handle a range of programming languages effectively.

After tokenization, each token $t_i$ is mapped to a high-dimensional vector embedding $\mathbf{v}_i \in \mathbb{R}^d$, where $d$ is the dimensionality of the embedding space. These embeddings are pre-trained using large datasets and encode valuable information about each token. In GraphCodeBERT, including data flow information enriches these embeddings, allowing for a representation that captures both the lexical and structural aspects of the code.

\subsection{Code Similarity Detection}
Code similarity detection is essential in software engineering for tasks like clone identification \cite{white2016deep} and automated code review \cite{key-martinez-ijseke}. Earlier methods relied on syntactic matching or basic structural analysis, which often missed deeper semantic relationships between code fragments \cite{key-martinez-swqd}.

More recent methods use pre-trained models to measure code similarity through learned embeddings. These models, such as GraphCodeBERT, compute similarity scores between code fragments using embeddings, capturing more subtle differences and similarities than syntax-based approaches can \cite{key-martinez-codesim}.

Advanced techniques also use attention mechanisms to focus on the most relevant parts of the code during similarity computation \cite{martinezgil2024improvingsourcecodesimilarity}. This helps the models assign importance to tokens based on their role in the overall semantic structure, improving the accuracy of similarity assessments. 

\subsection{Visualization Techniques for Code Comparison}
Visualization is an effective way to interpret the results of code similarity analysis \cite{smilkov2016embedding}. Current methods often present similarity matrices or heatmaps, where the degree of similarity between tokens from different code fragments is represented with color gradients. This offers an intuitive way to explore relationships between code segments and allows for detailed analysis by emphasizing areas of greater similarity.

GraphCodeBERT’s integration of structural information from data flow graphs would improve the clarity of these visualizations. It would allow users to see which tokens are similar and how data moves through the code. This is useful for tasks where understanding the relationships between code elements is crucial.

\subsection{Contribution over the State-of-the-Art}
Earlier approaches have face the interpretability challenge using less advanced models \cite{Abid2023}. This work proposes a method that improves transparency in code similarity assessments by utilizing GraphCodeBERT to identify relationships between code fragments. This method can detect code clones and explains why two code pieces are considered similar by revealing their connections. These explanations allow developers to understand the matching process better and gain more confidence in the results.

\section{Methodology}
Our approach uses the pre-trained GraphCodeBERT transformer model to capture deep semantic relationships within code, providing an improved understanding beyond mere syntactical analysis. The methodology is divided into several key steps: starting with the use of the pre-trained GraphCodeBERT model, we proceed through the processes of tokenization, input representation, and the application of an attention mechanism to determine the relative significance of each token within the code. Finally, we describe the result generation, which is visualized in diverse ways, offering an intuitive graphical representation of the semantic connections between tokens in different code fragments. 

\subsection{Pre-trained GraphCodeBERT Model}

GraphCodeBERT has been specifically fine-tuned on large-scale code datasets across multiple programming languages. This model captures complex patterns in code, making it ideal for tasks needing deep code understanding. Mathematically, the model can be expressed as a function $\mathcal{F}: \mathcal{T} \rightarrow \mathbb{R}^d$, where $\mathcal{T} = \{t_1, t_2, \dots, t_n\}$ is the input tokens, and $\mathcal{V} = \{\mathbf{v}_1, \mathbf{v}_2, \dots, \mathbf{v}_n\}$ is the corresponding vector embeddings. Formally:

\begin{equation}
\mathcal{V} = \mathcal{F}(\mathcal{T}) = \{\mathbf{v}_1, \mathbf{v}_2, \dots, \mathbf{v}_n\}, \quad \mathbf{v}_i \in \mathbb{R}^d
\end{equation}

\subsection{Tokenization}
Tokenization splits the source code into meaningful units called tokens (e.g., keywords, operators, or identifiers). These tokens are then transformed into high-dimensional vector embeddings that encode semantic information.

The process of tokenizing a given code fragment $c$ is performed by splitting it into a sequence of tokens $T = \{t_1, t_2, \dots, t_n\}$. The tokenization can be formalized as:

\begin{equation}
T = \text{Tokenize}(c) = \{t_i \mid t_i \in \text{Tokens}(c)\}
\end{equation}

where each token $t_i$ is identified based on alphanumeric characters or operators that are crucial to the code's syntax and structure. The output sequence $T$ serves as the input to the embedding layer.

\subsection{Input Representation}

Each token $t_i$ is mapped to a vector embedding $\mathbf{v}_i$ through the pre-trained GraphCodeBERT model:

\begin{equation}
\mathbf{v}_i = \mathcal{F}(t_i), \quad \mathbf{v}_i \in \mathbb{R}^d
\end{equation}

The input sequence $\mathcal{V}$ is then represented as:

\begin{equation}
\mathcal{V} = \{\mathbf{v}_1, \mathbf{v}_2, \dots, \mathbf{v}_n\}
\end{equation}

\subsection{Attention Mechanism}

The attention mechanism in GraphCodeBERT calculates attention weights for each token. These weights indicate the significance of one token to others in the sequence. This helps the model focus on the most relevant tokens when assessing the similarity between code fragments. Let $\mathbf{Q}$ and $\mathbf{K}$ represent the query and key matrices, respectively, derived from the input embeddings:

\begin{equation}
\mathbf{Q} = \mathbf{W}_Q \mathcal{V}, \quad \mathbf{K} = \mathbf{W}_K \mathcal{V}, \quad \mathbf{V} = \mathbf{W}_V \mathcal{V}
\end{equation}

where $\mathbf{W}_Q, \mathbf{W}_K, \mathbf{W}_V \in \mathbb{R}^{d \times d}$ are learnable weight matrices. The scaled dot-product attention is computed as:

\begin{equation}
\mathbf{A} = \text{softmax}\left(\frac{\mathbf{Q} \mathbf{K}^\top}{\sqrt{d_k}}\right) \mathbf{V}
\end{equation}

where $d_k$ is the dimensionality of the key vectors. The resulting attention matrix $\mathbf{A} \in \mathbb{R}^{n \times n}$ represents the weight assigned to each token pair, indicating the influence of token $t_j$ on token $t_i$.

\subsection{Similarity Matrix}

Given two code fragments $c_1$ and $c_2$, with their respective token sequences $T_1 = \{t_1^{(1)}, t_2^{(1)}, \dots, t_{n_1}^{(1)}\}$ and $T_2 = \{t_1^{(2)}, t_2^{(2)}, \dots, t_{n_2}^{(2)}\}$, and their corresponding attention matrices $\mathbf{A}_1$ and $\mathbf{A}_2$, the computation of the similarity matrix $\mathbf{S}$ should be:

\begin{equation}
\mathbf{S} = \mathbf{A}_1 \mathbf{A}_2^\top \in \mathbb{R}^{n_1 \times n_2}
\end{equation}

The element $S_{ij}$ in the similarity matrix represents the semantic similarity between token $t_i^{(1)}$ from $T_1$ and token $t_j^{(2)}$ from $T_2$. In order to capture higher-order relationships, the similarity matrix can be refined through multiple attention heads:

\begin{equation}
\mathbf{S} = \frac{1}{H} \sum_{h=1}^H \mathbf{A}_1^{(h)} \mathbf{A}_2^{(h)\top}
\end{equation}

where $H$ is the number of attention heads.

\subsection{Dimensionality Reduction and Visualization}
In order to effectively visualize the high-dimensional embeddings generated by GraphCodeBERT, we use Principal Component Analysis (PCA) \cite{Musil19}, t-distributed Stochastic Neighbor Embedding (t-SNE) \cite{abs-2306-11898}, and Uniform Manifold Approximation and Projection (UMAP) \cite{abs-2306-11898}. Each of these methods serves a distinct purpose in the dimensionality reduction process.

PCA reduces the dimensionality while preserving the data structure in terms of variance. In fact, it can identify the principal components (linear combinations of the original variables) that capture the maximum variance in the data. PCA is defined as:

\begin{equation} \mathbf{V}{\text{PCA}} = \mathbf{V} \mathbf{W}{\text{PCA}}, \quad \mathbf{W}_{\text{PCA}} \in \mathbb{R}^{d \times p} \end{equation}

where $p < d$ is the reduced dimensionality, $\mathbf{V}$ represents the original data, and $\mathbf{W}_{\text{PCA}}$ contains the principal component vectors. PCA is a good method for initial exploration.

t-SNE focuses on preserving the local structure of the data, making it well-suited for visualizing clusters and relationships. t-SNE maps high-dimensional points $\mathbf{v}_i$ to lower-dimensional points $\mathbf{y}_i \in \mathbb{R}^q$ by minimizing the Kullback-Leibler divergence between the joint probabilities of the high-dimensional and low-dimensional points:

\begin{equation} \text{KL}(P \parallel Q) = \sum_{i \neq j} P_{ij} \log \frac{P_{ij}}{Q_{ij}} \end{equation}

where $P_{ij}$ and $Q_{ij}$ are the joint probabilities of pairs of points in the high-dimensional and low-dimensional spaces, respectively. t-SNE is effective for visualizing small-scale patterns, but it can be computationally intensive and may struggle with huge datasets.

UMAP is a more recent technique that combines the strengths of both PCA and t-SNE while addressing some limitations. UMAP aims to preserve the data's local and global structures. To do that, it builds a high-dimensional graph representation of the data, which optimizes for a low-dimensional layout. 

This approach is formalized as an optimization problem where a topological structure represents the high-dimensional graph, and the low-dimensional embedding is optimized to preserve this structure.

The idea is to build a weighted graph \( G = (V, E) \) where \( V \) are the vertices representing data points and \( E \) are the edges connecting these points. The edge weights are determined by pairwise similarities between the points and are calculated using the formula:

\[
w_{ij} = \exp\left(-\frac{d(v_i, v_j) - \rho_i}{\sigma_i}\right)
\]

where:
\begin{itemize}
  \item \( d(v_i, v_j) \) is the distance between points \( v_i \) and \( v_j \) in the original high-dimensional space.
  \item \( \rho_i \) is the distance to the nearest neighbor of \( v_i \).
  \item \( \sigma_i \) is a scaling factor controlling the distance distribution's spread.
\end{itemize}

UMAP then optimizes the low-dimensional embedding \( Y \) by minimizing a cross-entropy loss between the high-dimensional graph and the low-dimensional representation:

\[
\text{argmin}_{Y} \sum_{(i,j) \in E} w_{ij} \log \left(\frac{w_{ij}}{\text{dist}(y_i, y_j)}\right) + (1 - w_{ij}) \log \left(\frac{1 - w_{ij}}{1 - \text{dist}(y_i, y_j)}\right)
\]

where:
\begin{itemize}
  \item \( \text{dist}(y_i, y_j) \) is the distance between the low-dimensional points \( y_i \) and \( y_j \) in the embedding.
  \item \( Y \) represents the coordinates of all points in the low-dimensional space.
\end{itemize}

In order to make the analysis more accessible to developers, we incorporate a range of advanced data visualization techniques. These methods improve the clarity of relationships between fragments and allow us to explore the data more deeply. In addition, we use saliency maps to interpret the importance of individual tokens in the embeddings. Saliency maps help identify which tokens contribute the most to the similarity to generate a visual representation that shows the most influential tokens.  

Let the similarity score between two code fragments be denoted as $\text{sim}(\mathbf{F}_1, \mathbf{F}_2)$. The saliency map for a token embedding $\mathbf{e}_i^{(1)}$ from the first fragment is computed as the gradient of the similarity score concerning that embedding:

\[
\text{saliency}(\mathbf{e}_i^{(1)}) = \left\| \frac{\partial \text{sim}(\mathbf{F}_1, \mathbf{F}_2)}{\partial \mathbf{e}_i^{(1)}} \right\|
\]

Visualizing these saliency maps alongside the code fragments aims to help developers better understand which parts of the code are driving the similarity.

\subsection{Similarity Computation}
In order to quantify the similarity between two code fragments, their respective token embeddings are compared using cosine similarity. Let $\mathbf{e}_i^{(1)}$ and $\mathbf{e}_j^{(2)}$ denote the embeddings of the $i$-th token in the first fragment and the $j$-th token in the second fragment, respectively. The cosine similarity between these embeddings is computed as:

\[
\text{cosine\_sim}(\mathbf{e}_i^{(1)}, \mathbf{e}_j^{(2)}) = \frac{\mathbf{e}_i^{(1)} \cdot \mathbf{e}_j^{(2)}}{\|\mathbf{e}_i^{(1)}\| \|\mathbf{e}_j^{(2)}\|}
\]

The resulting  matrix $\mathbf{S}$, where $S_{ij} = \text{cosine\_sim}(\mathbf{e}_i^{(1)}, \mathbf{e}_j^{(2)})$, serves as the basis for further analysis.

\subsection{Interpretation Strategy}
Our strategy for augmenting interpretability involves multiple visualization methods to process the embeddings and their similarities:

\begin{itemize}
	\item PCA and t-SNE are used to reduce the dimensionality of the embeddings, allowing for visual inspection of the embedding space. These methods reveal clusters in the data.

	\item UMAP is used to visualize high-dimensional data. UMAP preserves more of the global structure of the data compared to t-SNE and can often produce more meaningful clusters, making it particularly useful for exploring relationships and groupings within the embeddings.

	\item Saliency maps are used to visualize the contribution of individual features in the embedding to the output. The idea is to identify the parts of the input that influence the model’s predictions most. 
\end{itemize}

\section{Use Case}
We introduce a method to improve understanding of how similarity between classical sorting algorithms can be evaluated. Sorting algorithms are fundamental to computer science and the foundation for many systems. This study focuses on five well-known sorting algorithms: 

\begin{itemize}
	\item \textbf{Bubble Sort}: A simple comparison-based algorithm that repeatedly swaps adjacent elements if they are in the wrong order, slowly bubbling the largest elements to the end.
	
	\item \textbf{Selection Sort}: An algorithm that repeatedly selects the smallest element from the unsorted part of the list and swaps it with the first unsorted element.
	
	\item \textbf{Insertion Sort}: A sorting algorithm that builds the final sorted list one element at a time by inserting each element into its correct position among the previously sorted elements.
	
	\item \textbf{Merge Sort}: A divide-and-conquer algorithm that splits the list into smaller sublists, recursively sorts them, and then merges the sorted sublists.
	
	\item \textbf{Quick Sort}: A divide-and-conquer algorithm that selects a pivot element, partitions the list around the pivot, and recursively sorts the sublists.
\end{itemize}

We provide a new perspective on how these algorithms relate to one another beyond their operational characteristics. The aim is to move beyond traditional analysis methods and focus on their code representations to identify aspects that may not be apparent through standard analysis.

\subsection*{Bubble Sort}
\begin{lstlisting}[language=Python, caption=An implementation of the Bubble Sort algorithm in Python. Bubble Sort is a simple comparison-based sorting algorithm that repeatedly steps through the list.]
def bubble_sort(arr):
    n = len(arr)
    for i in range(n):
        for j in range(0, n-i-1):
            if arr[j] > arr[j+1]:
                arr[j], arr[j+1] = arr[j+1], arr[j]
\end{lstlisting}

\subsection*{Selection Sort}
\begin{lstlisting}[language=Python, caption=An implementation of the Selection Sort algorithm in Python. Selection Sort is a straightforward comparison-based sorting algorithm that repeatedly selects the smallest element from the unsorted portion of the list.]
def selection_sort(arr):
    for i in range(len(arr)):
        min_idx = i
        for j in range(i+1, len(arr)):
            if arr[j] < arr[min_idx]:
                min_idx = j
        arr[i], arr[min_idx] = arr[min_idx], arr[i]
    return arr
\end{lstlisting}

\subsection*{Merge Sort}
\begin{lstlisting}[language=Python, caption=An implementation of the Merge Sort algorithm in Python that recursively divides the array into two halves sorts each half and then merges them.]
def merge_sort(arr):
    if len(arr) > 1:
        mid = len(arr) // 2
        L = arr[:mid]
        R = arr[mid:]

        merge_sort(L)
        merge_sort(R)

        i = j = k = 0
        while i < len(L) and j < len(R):
            if L[i] < R[j]:
                arr[k] = L[i]
                i += 1
            else:
                arr[k] = R[j]
                j += 1
            k += 1

        while i < len(L):
            arr[k] = L[i]
            i += 1
            k += 1

        while j < len(R):
            arr[k] = R[j]
            j += 1
            k += 1
\end{lstlisting}

\subsection*{Quick Sort}
\begin{lstlisting}[language=Python, caption=An implementation of the Quick Sort algorithm in Python that selects a pivot \, partitions the array around the pivot and recursively sorts the subarrays.]
def partition(arr, low, high):
    pivot = arr[high]
    i = low - 1

    for j in range(low, high):
        if arr[j] <= pivot:
            i = i + 1
            arr[i], arr[j] = arr[j], arr[i]

    arr[i + 1], arr[high] = arr[high], arr[i + 1]
    return i + 1

def quick_sort(arr, low, high):
    if low < high:
        pi = partition(arr, low, high)

        quick_sort(arr, low, pi - 1)
        quick_sort(arr, pi + 1, high)
\end{lstlisting}

\subsection*{Insertion Sort}
\begin{lstlisting}[language=Python, caption=An implementation of the Insertion Sort algorithm in Python that builds the sorted array one element at a time by repeatedly inserting elements.]
def insertion_sort(arr):
    for i in range(1, len(arr)):
        key = arr[i]
        j = i-1
        while j >=0 and key < arr[j]:
            arr[j + 1] = arr[j]
            j -= 1
        arr[j + 1] = key
    return arr
\end{lstlisting}

\begin{figure}
	\centering
	\includegraphics[width=0.99\textwidth]{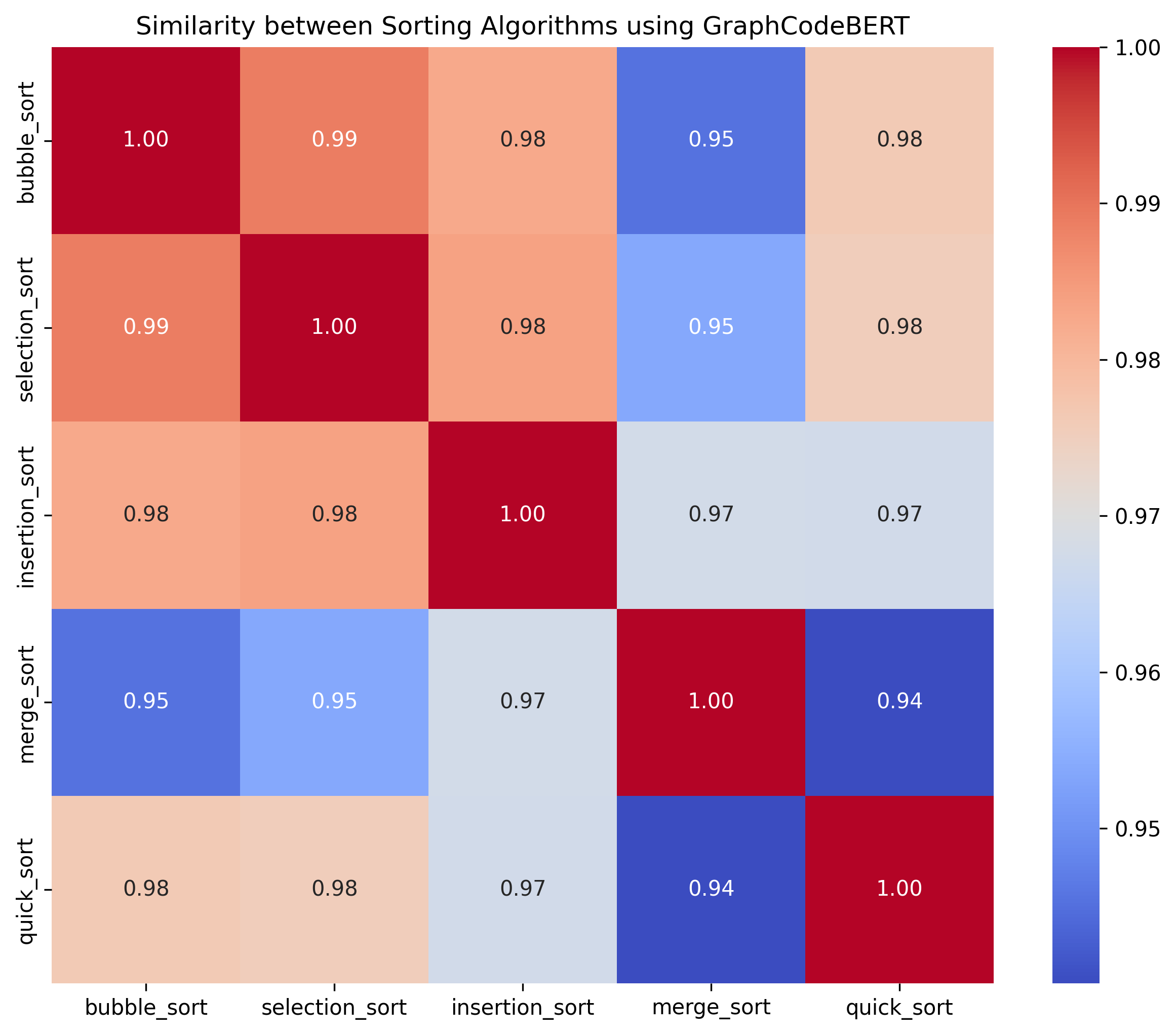}
	\caption{Visualization of the similarity relationships among various classical sorting algorithms using GraphCodeBERT. This heatmap shows how similar sorting algorithms are, based on their structure and behavior, using GraphCodeBERT. Algorithms like Bubble Sort and Insertion Sort, which follow similar step-by-step processes, show higher similarity scores. On the other hand, Quick Sort, which uses a more complex partitioning approach, shows lower similarity with Bubble Sort. }
	\label{fig:sorting_algorithms_similarity}
\end{figure}

Each algorithm was implemented in Python and represented as a code string. These code strings were tokenized and processed through GraphCodeBERT to generate vector embeddings, which encapsulate the structural and semantic properties of the code. The resulting similarities were visualized in a heatmap, as shown in Figure \ref{fig:sorting_algorithms_similarity}.

The heatmap reveals that algorithms sharing similar structural patterns or operational steps, such as Insertion Sort and Bubble Sort, exhibit higher similarity scores. At the same time, algorithms with fundamentally different approaches, like Quick Sort and Bubble Sort, display lower similarity scores. This approach provides interesting insights into different sorting algorithms' inherent structural and functional relationships. Our goal is that our approach can help us understand these results in greater depth.

\subsection{Pairwise Comparisons of Sorting Algorithms using PCA}
We present pairwise comparisons of classical sorting algorithms by projecting their token embeddings into a two-dimensional space using PCA. This method helps to visualize the relationships between the embeddings in a simplified form. Each image provided shows the distribution of token embeddings for a specific pair of sorting algorithms, with different colors assigned to each algorithm for a clear distinction. These visualizations allow us to observe how the embeddings of different algorithms group or spread out in the 2D space.

\begin{figure}[htbp]
    \centering

    \begin{subfigure}{0.48\textwidth}
        \includegraphics[width=\textwidth]{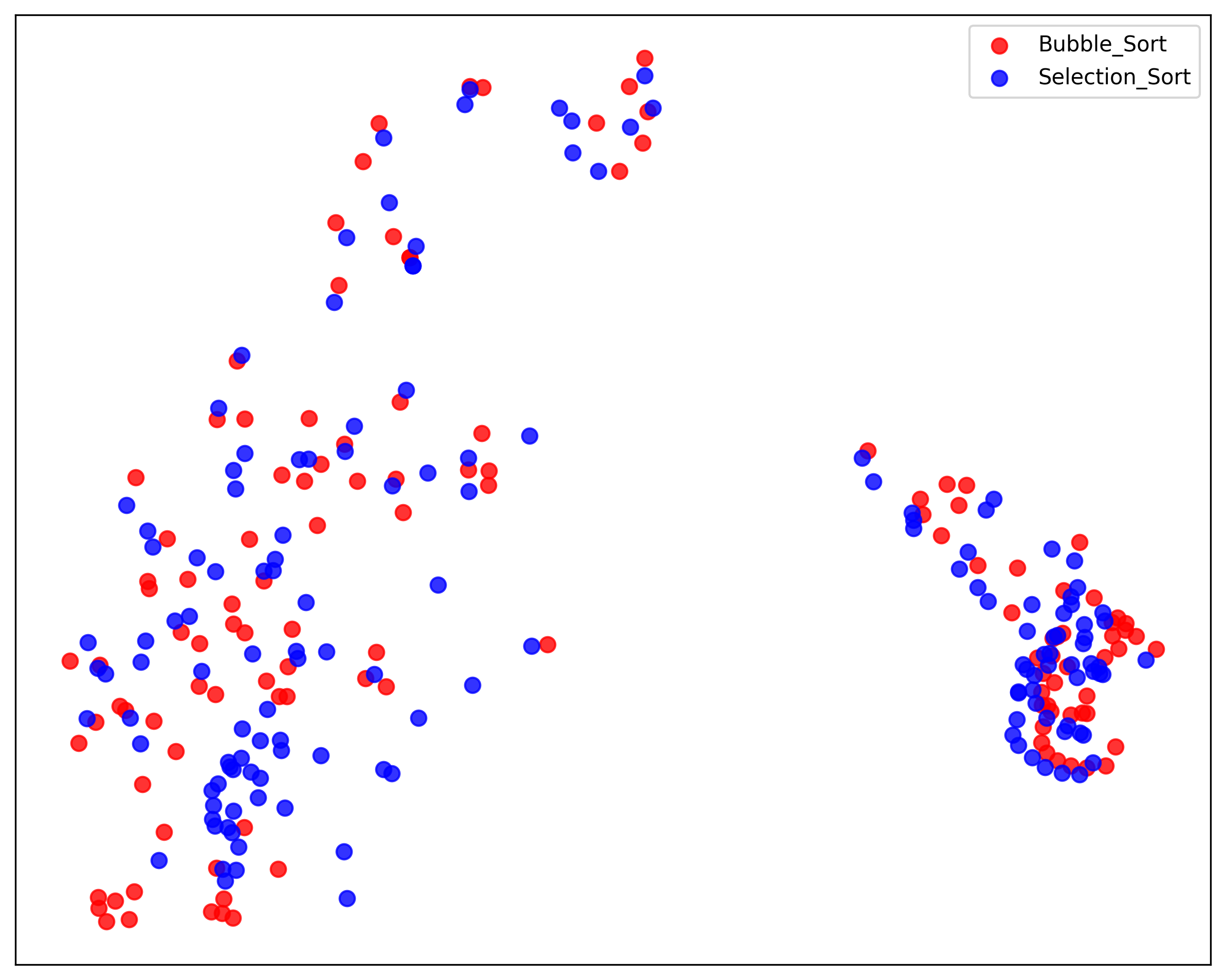}
        \caption{Bubble Sort vs. Selection Sort}
    \end{subfigure}
    \hfill
    \begin{subfigure}{0.48\textwidth}
        \includegraphics[width=\textwidth]{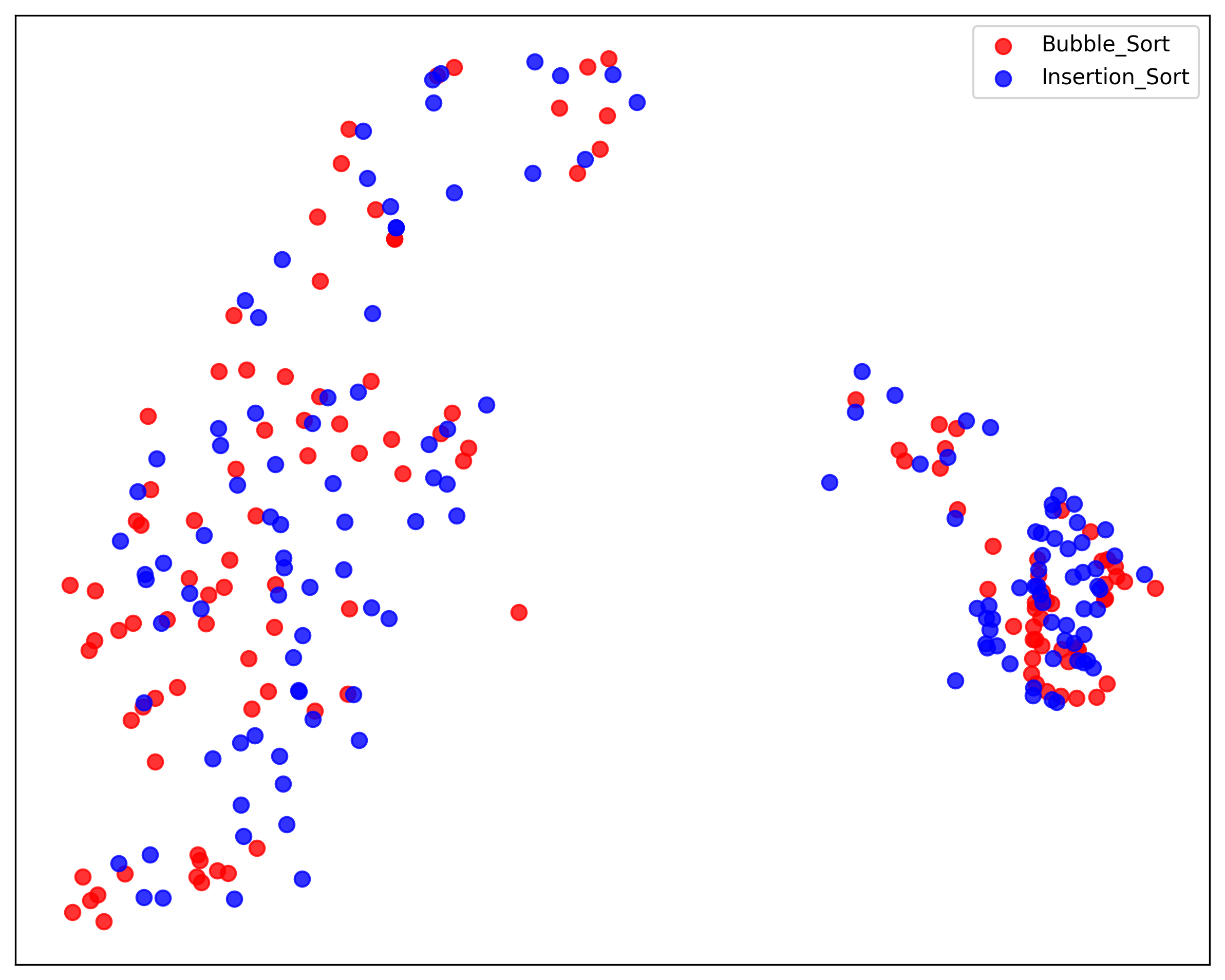}
        \caption{Bubble Sort vs. Insertion Sort}
    \end{subfigure}

    \vspace{0.3cm}

    \begin{subfigure}{0.48\textwidth}
        \includegraphics[width=\textwidth]{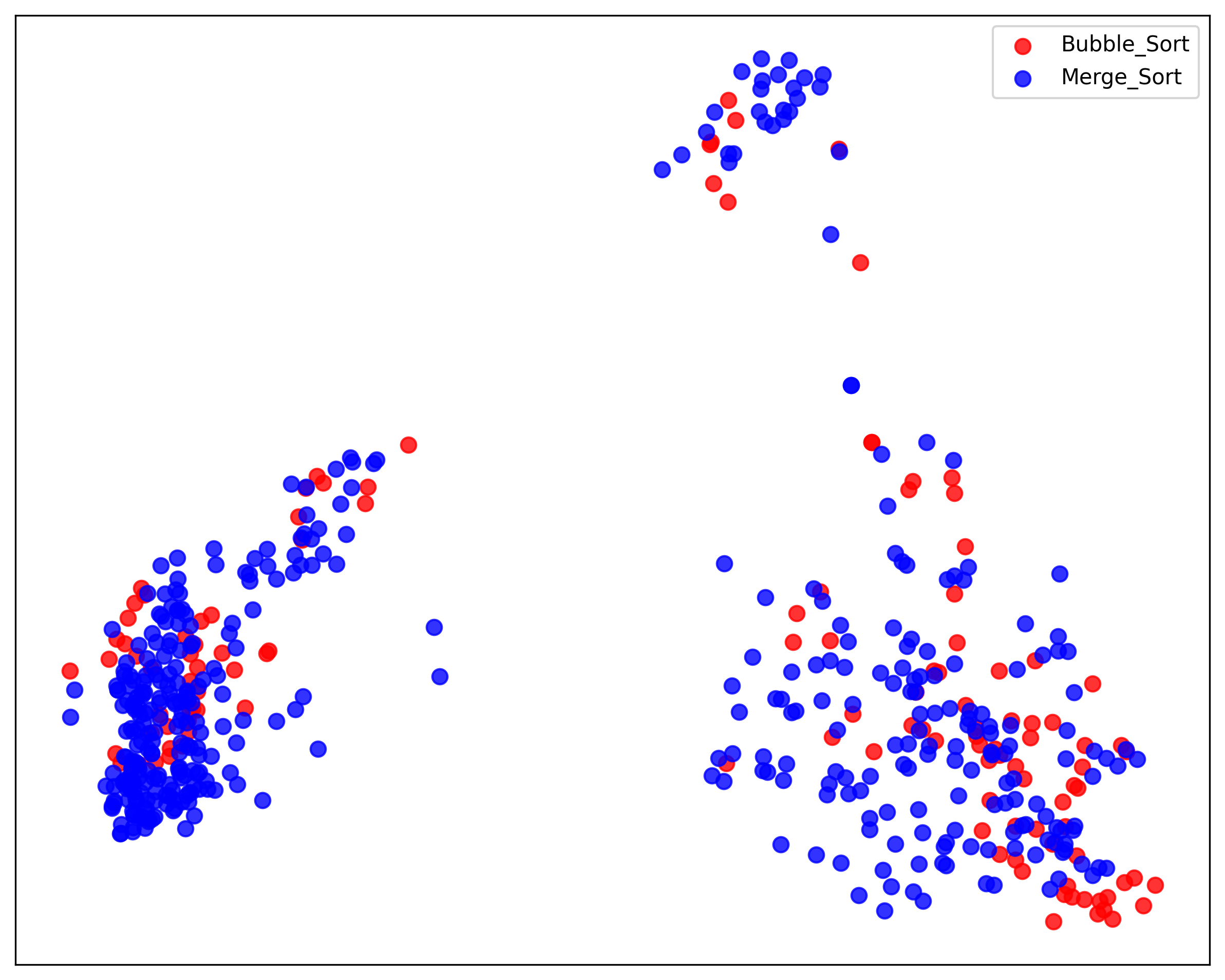}
        \caption{Bubble Sort vs. Merge Sort}
    \end{subfigure}
    \hfill
    \begin{subfigure}{0.48\textwidth}
        \includegraphics[width=\textwidth]{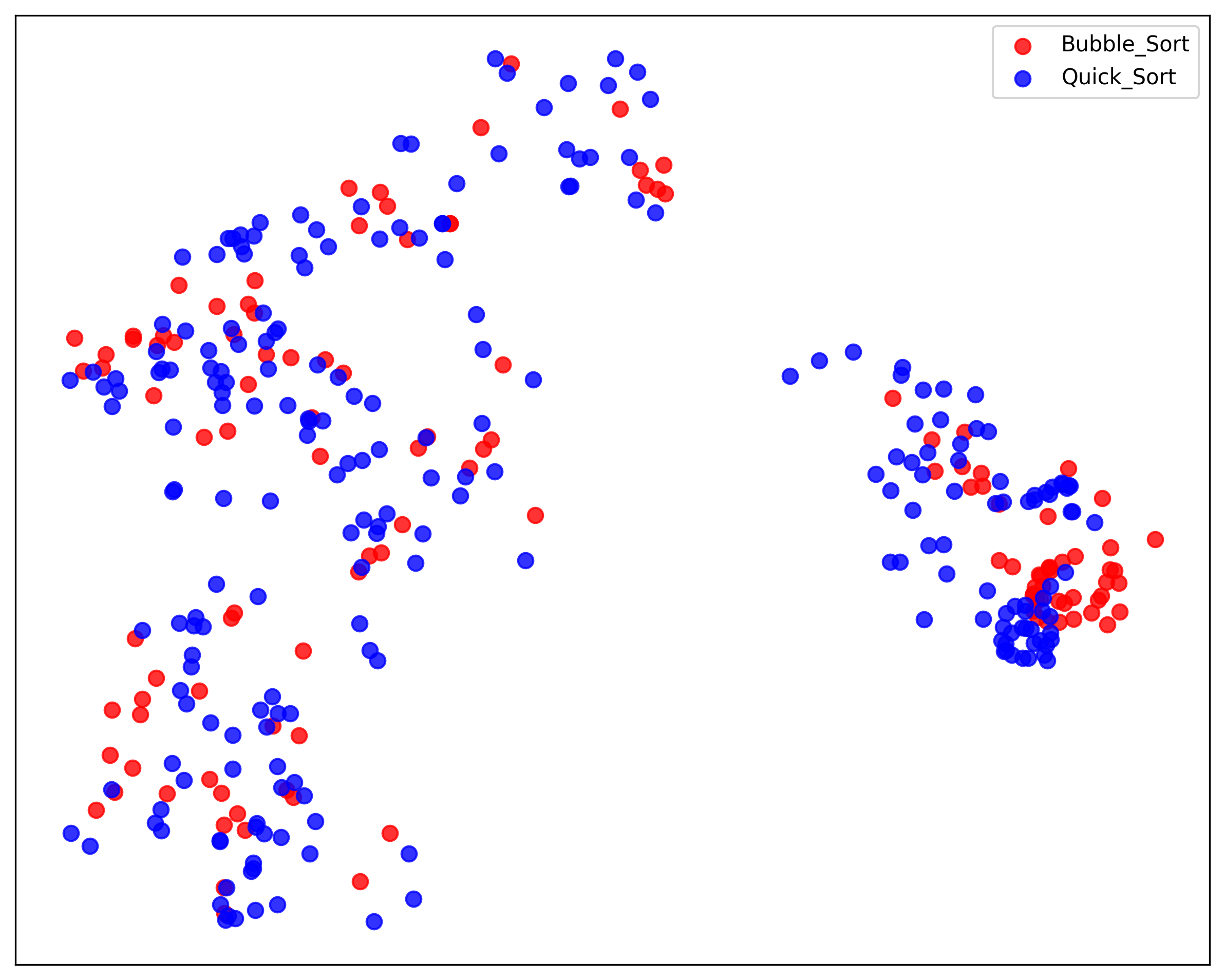}
        \caption{Bubble Sort vs. Quick Sort}
    \end{subfigure}

    \vspace{0.3cm}

    \begin{subfigure}{0.48\textwidth}
        \includegraphics[width=\textwidth]{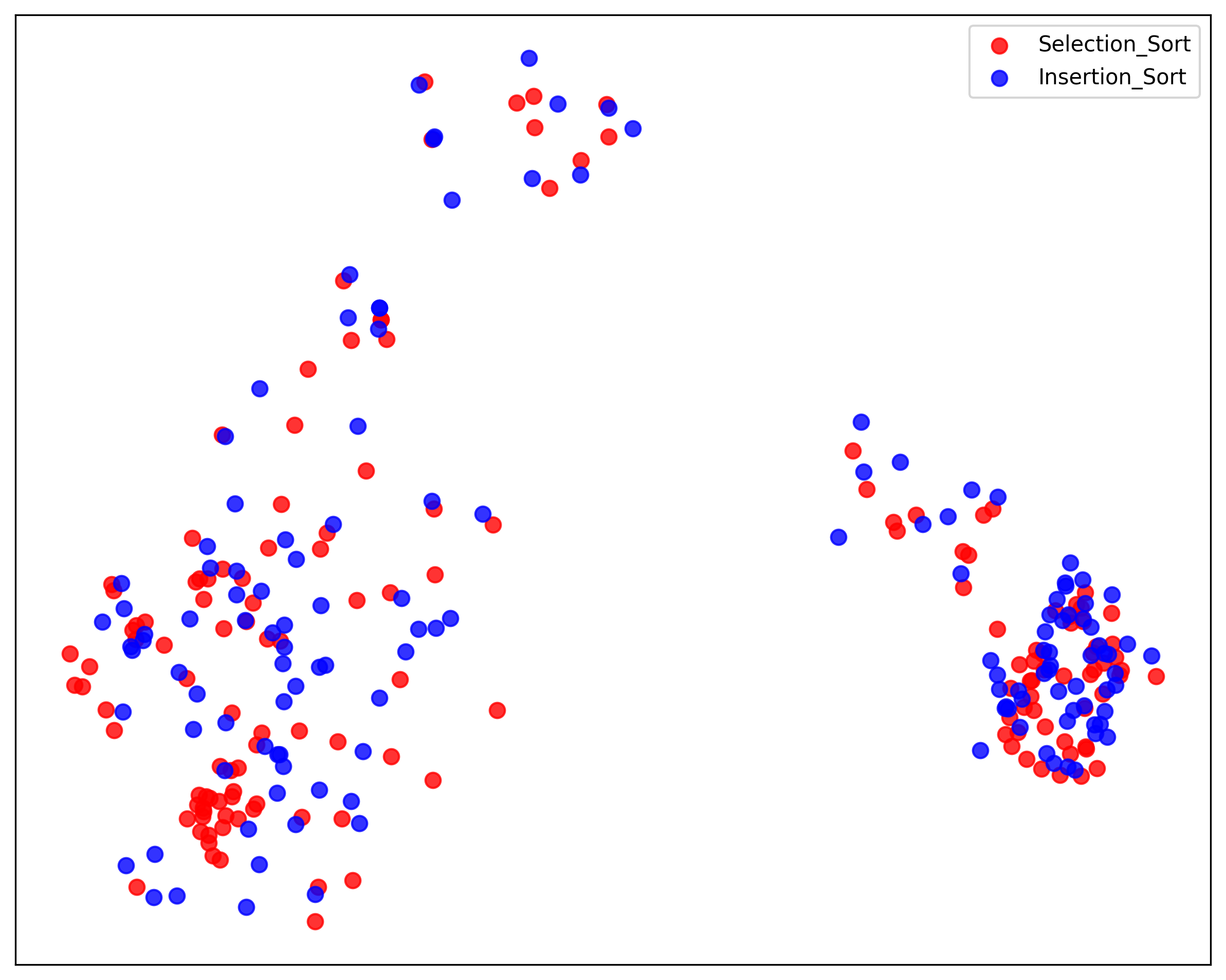}
        \caption{Selection Sort vs. Insertion Sort}
    \end{subfigure}

    \caption{Pairwise comparisons of classical sorting algorithms using PCA, showing the token embeddings in a 2D space (Part 1).}
    \label{fig:sorting-algorithms-comparisons-part1}
\end{figure}

\begin{figure}[htbp]
    \centering

    \begin{subfigure}{0.48\textwidth}
        \includegraphics[width=\textwidth]{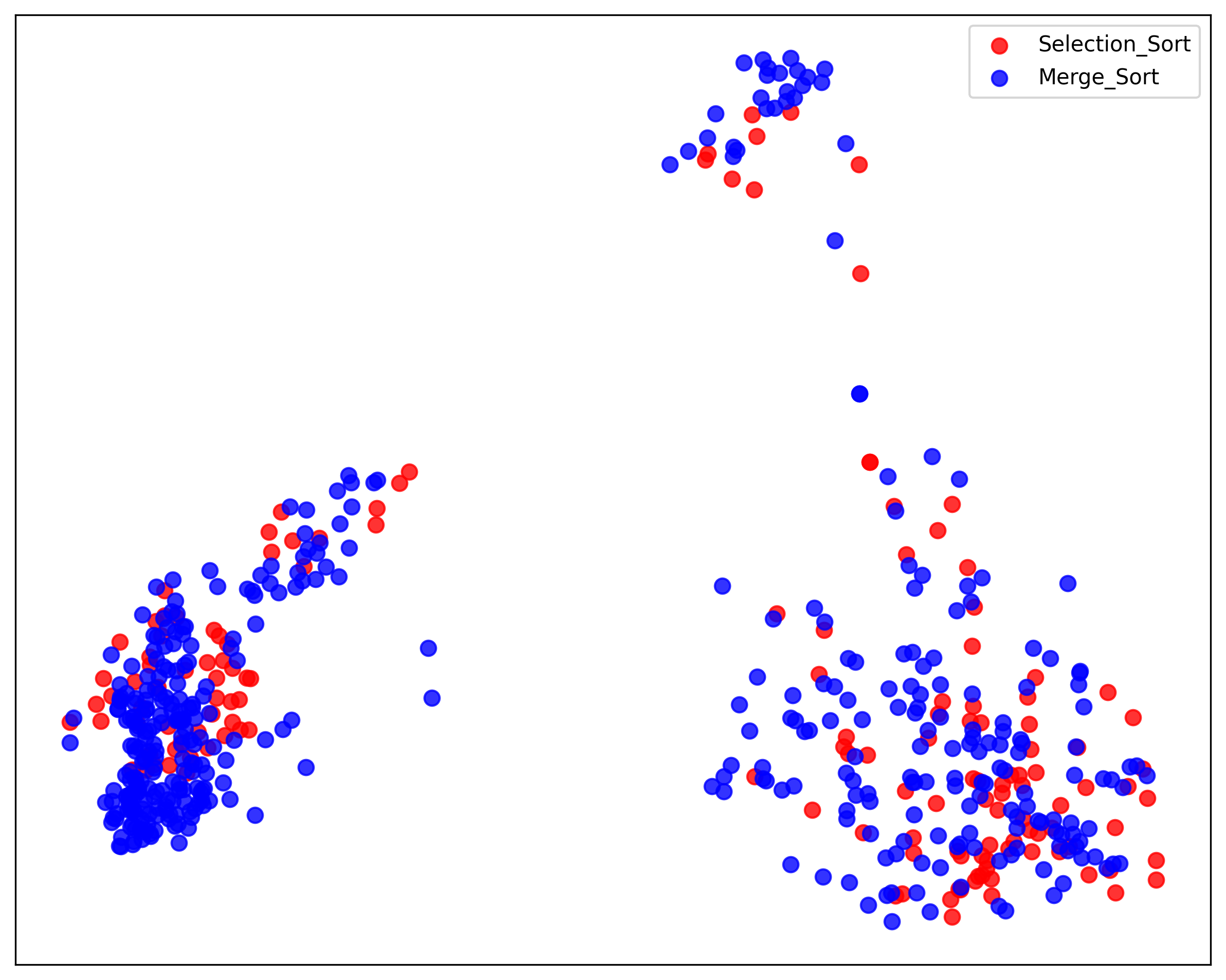}
        \caption{Selection Sort vs. Merge Sort}
    \end{subfigure}
    \hfill
    \begin{subfigure}{0.48\textwidth}
        \includegraphics[width=\textwidth]{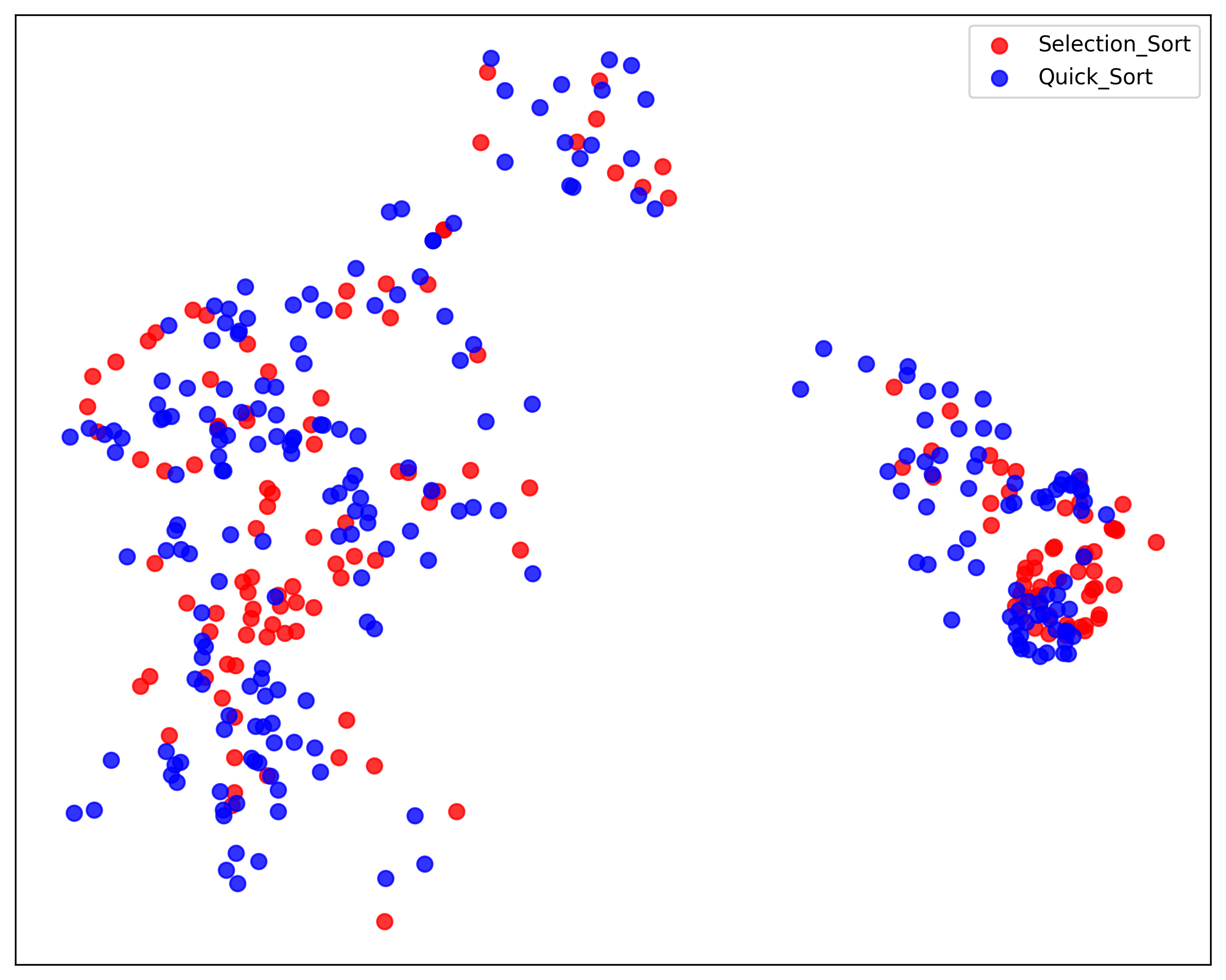}
        \caption{Selection Sort vs. Quick Sort}
    \end{subfigure}

    \vspace{0.3cm}

    \begin{subfigure}{0.48\textwidth}
        \includegraphics[width=\textwidth]{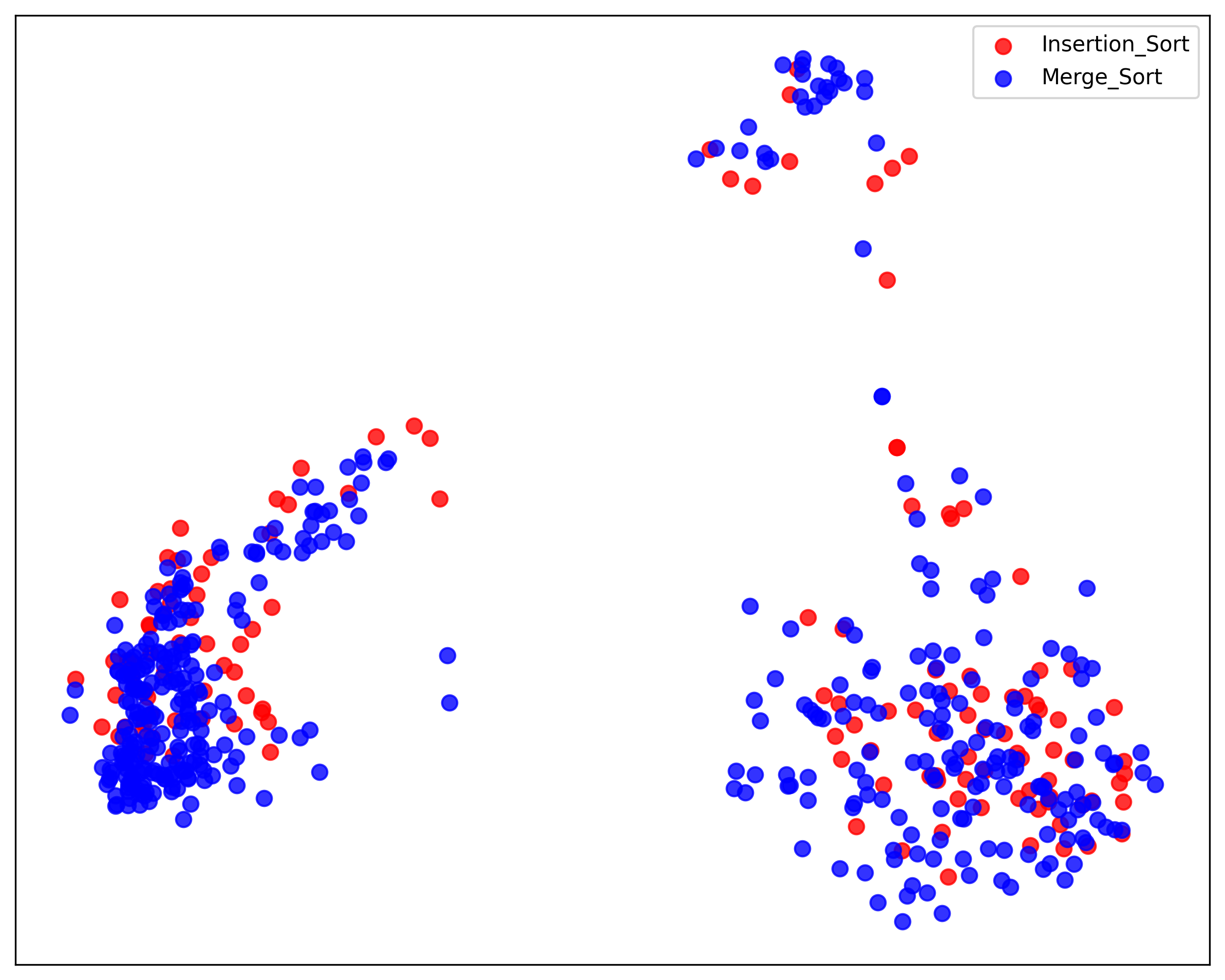}
        \caption{Insertion Sort vs. Merge Sort}
    \end{subfigure}
    \hfill
    \begin{subfigure}{0.48\textwidth}
        \includegraphics[width=\textwidth]{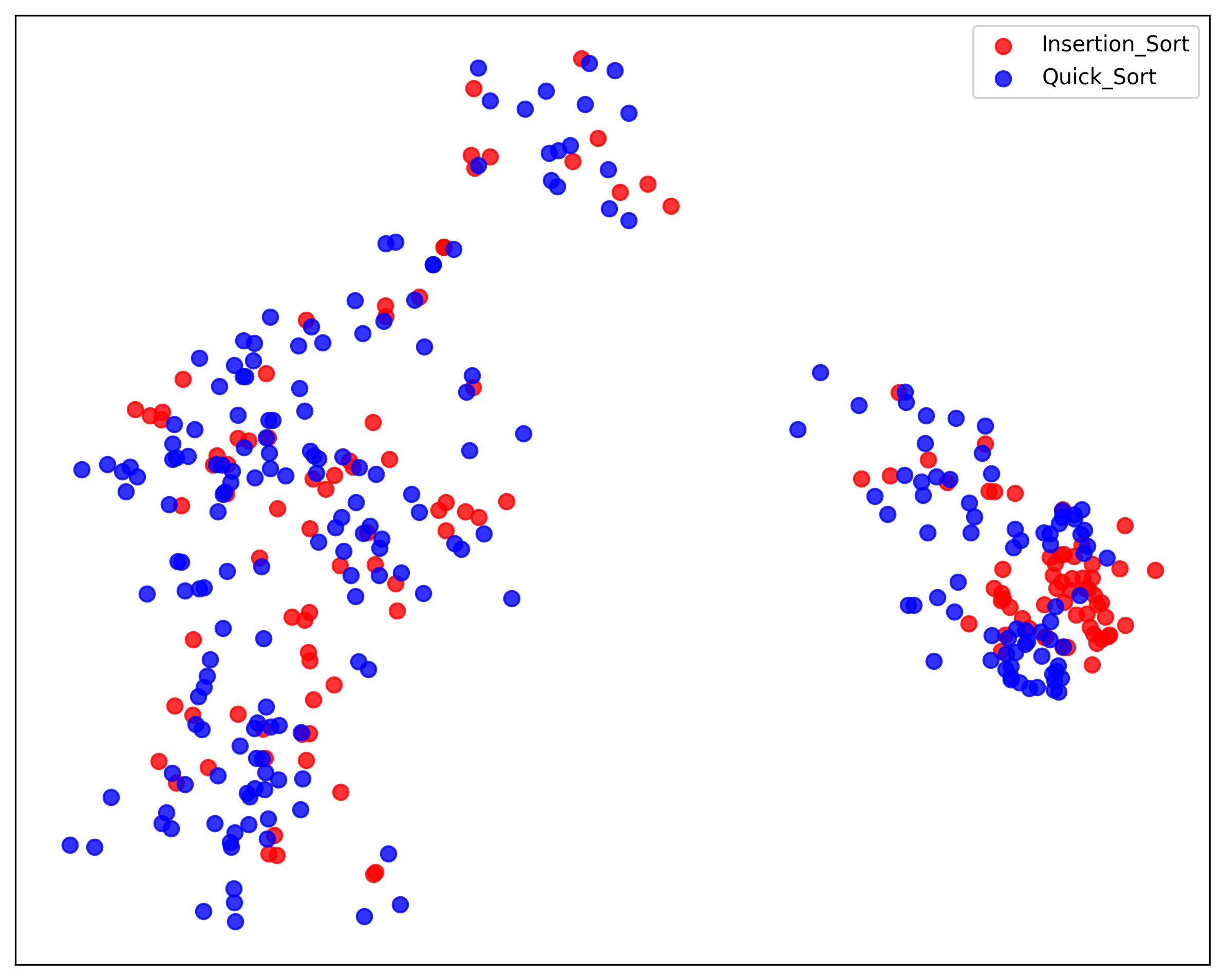}
        \caption{Insertion Sort vs. Quick Sort}
    \end{subfigure}

    \vspace{0.3cm}

    \begin{subfigure}{0.48\textwidth}
        \includegraphics[width=\textwidth]{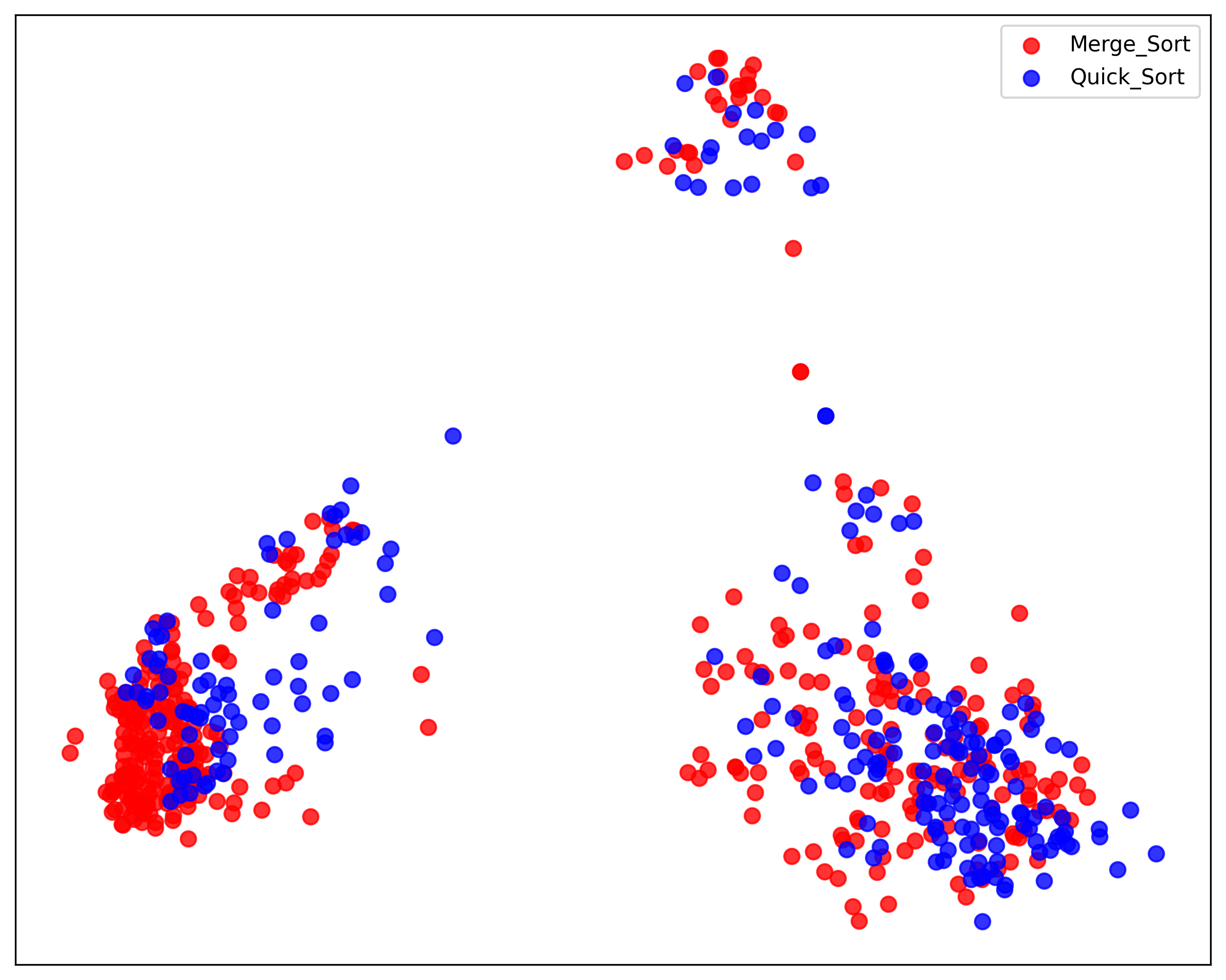}
        \caption{Merge Sort vs. Quick Sort}
    \end{subfigure}

    \caption{Pairwise comparisons of classical sorting algorithms using PCA, showing the token embeddings in a 2D space (Part 2).}
    \label{fig:sorting-algorithms-comparisons-part2}
\end{figure}

Figures \ref{fig:sorting-algorithms-comparisons-part1} and \ref{fig:sorting-algorithms-comparisons-part2} illustrate these pairwise comparisons across the mentioned sorting algorithms, visually representing their token embeddings.

\subsection{Pairwise Comparisons of Sorting Algorithms using t-SNE}
We also present the pairwise comparisons between classical sorting algorithms by visualizing their token embeddings in a two-dimensional space using t-SNE. This technique is particularly effective for high-dimensional data reduction, allowing us to project the token embeddings into a 2D space while preserving the local structure of the data. Once again, each image visually represents the token embeddings for a specific pair of sorting algorithms, with different colors assigned to each algorithm for better differentiation. These visualizations offer a more detailed understanding of how the token embeddings for the algorithms relate to each other.

\begin{figure}[htbp]
    \centering

    \begin{subfigure}{0.48\textwidth}
        \includegraphics[width=\textwidth]{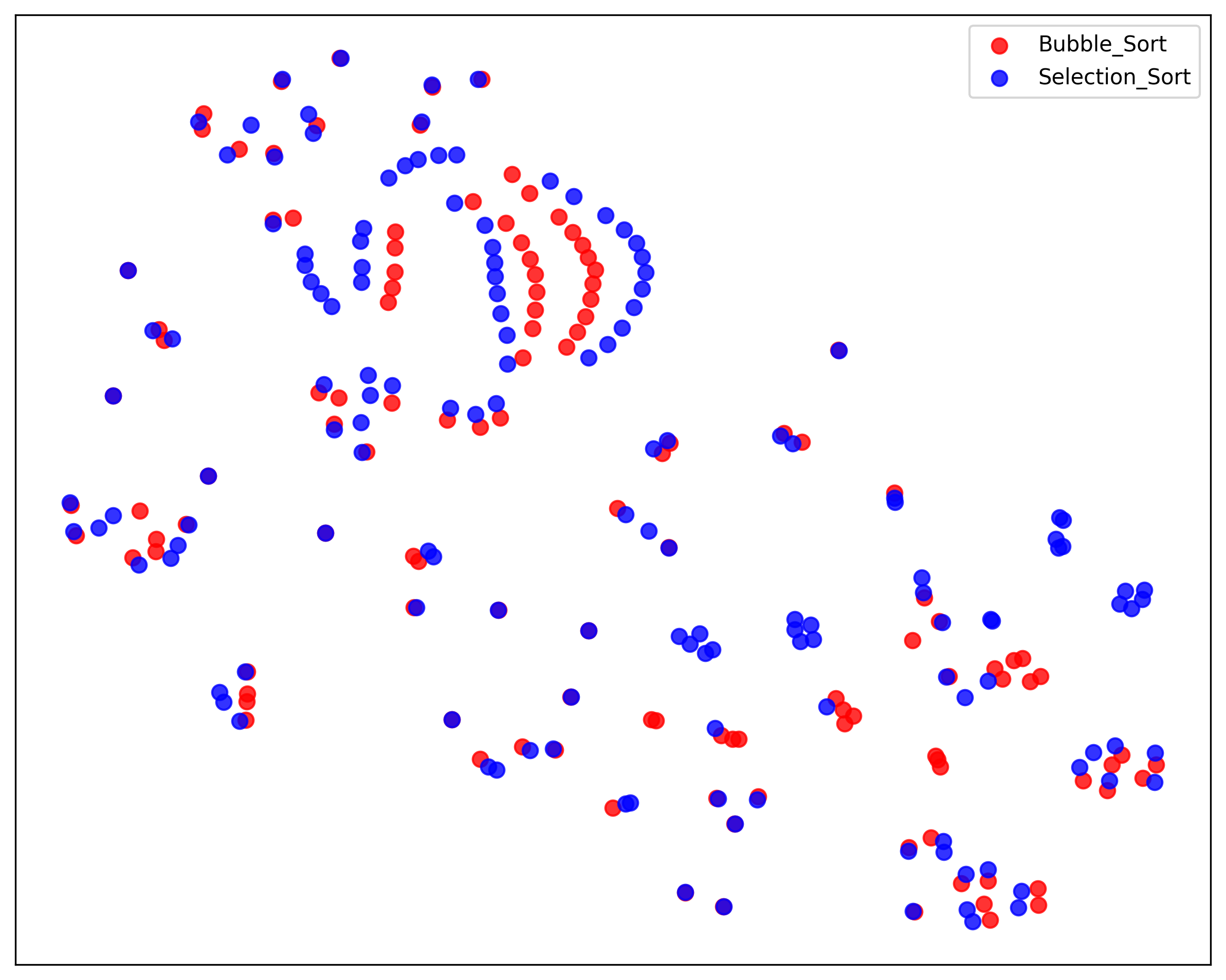}
        \caption{Bubble Sort vs. Selection Sort}
    \end{subfigure}
    \hfill
    \begin{subfigure}{0.48\textwidth}
        \includegraphics[width=\textwidth]{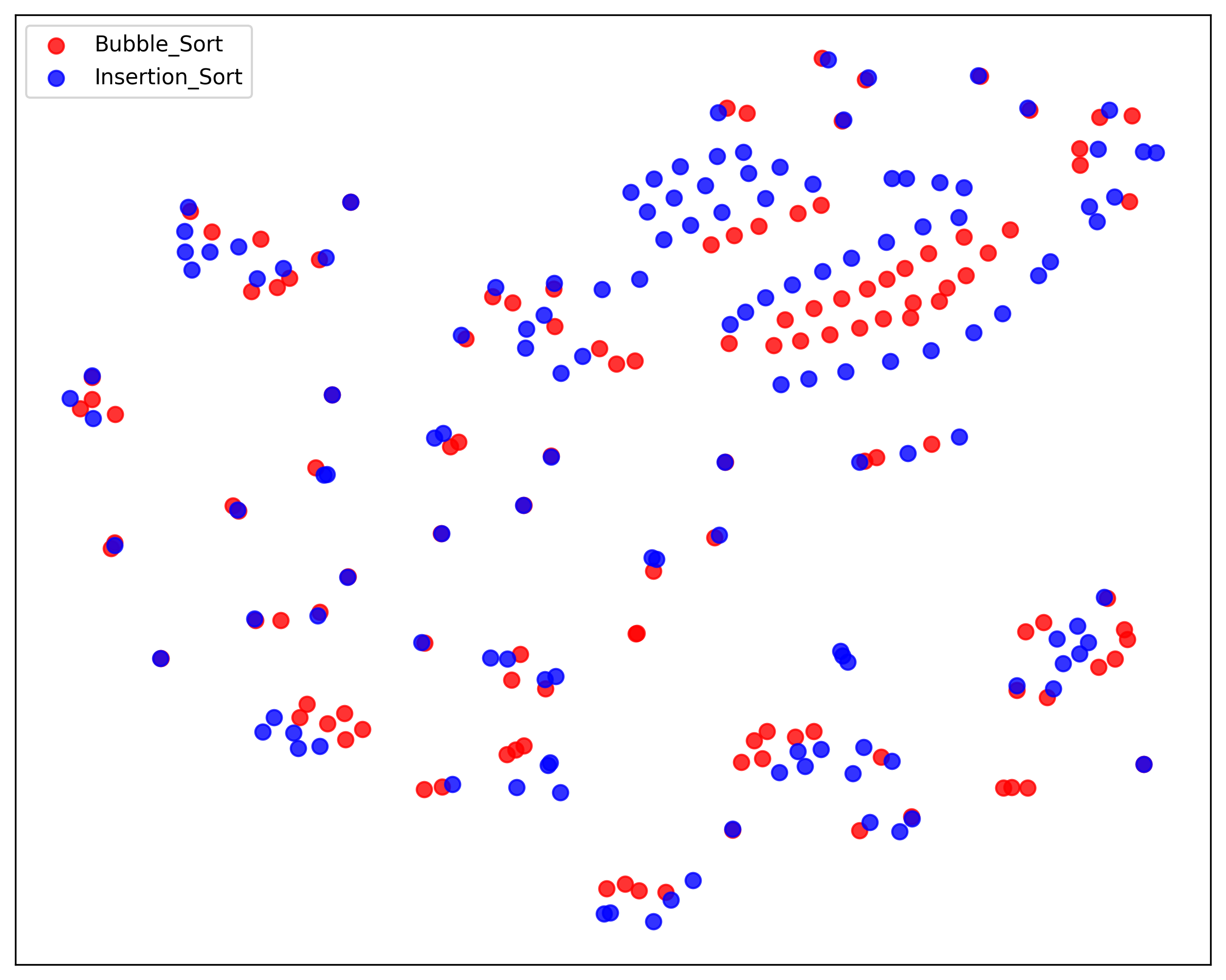}
        \caption{Bubble Sort vs. Insertion Sort}
    \end{subfigure}

    \vspace{0.3cm}

    \begin{subfigure}{0.48\textwidth}
        \includegraphics[width=\textwidth]{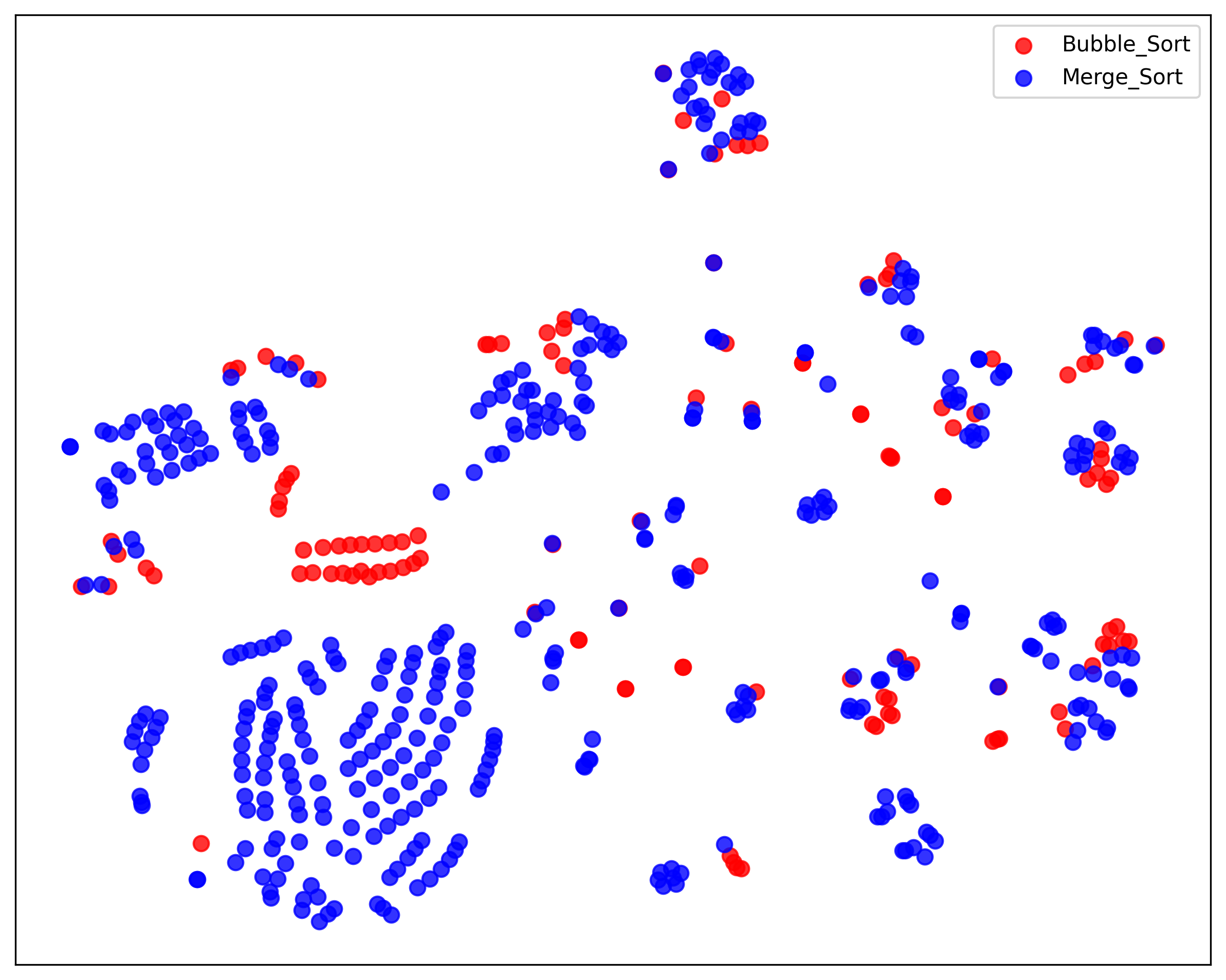}
        \caption{Bubble Sort vs. Merge Sort}
    \end{subfigure}
    \hfill
    \begin{subfigure}{0.48\textwidth}
        \includegraphics[width=\textwidth]{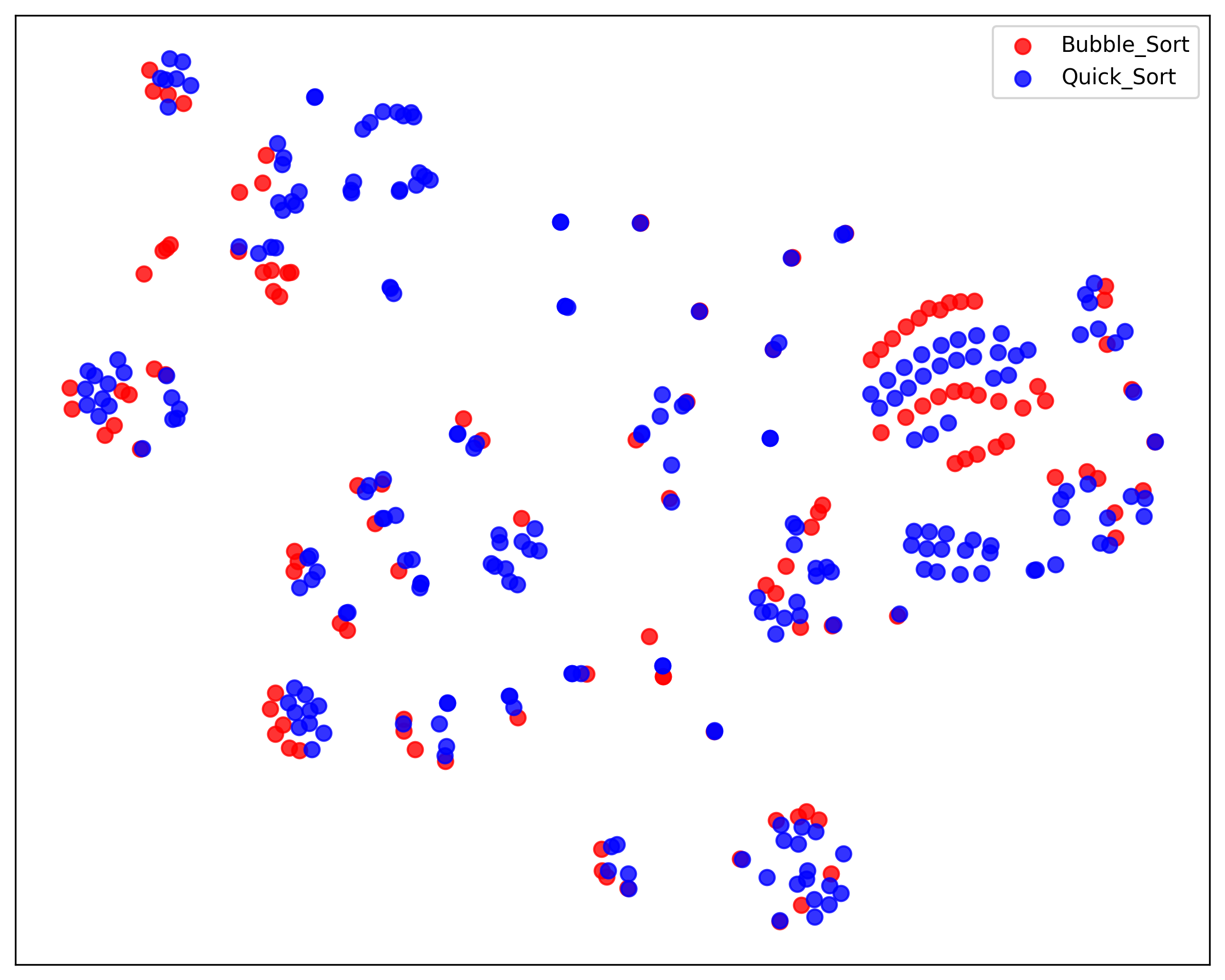}
        \caption{Bubble Sort vs. Quick Sort}
    \end{subfigure}

    \vspace{0.3cm}

    \begin{subfigure}{0.48\textwidth}
        \includegraphics[width=\textwidth]{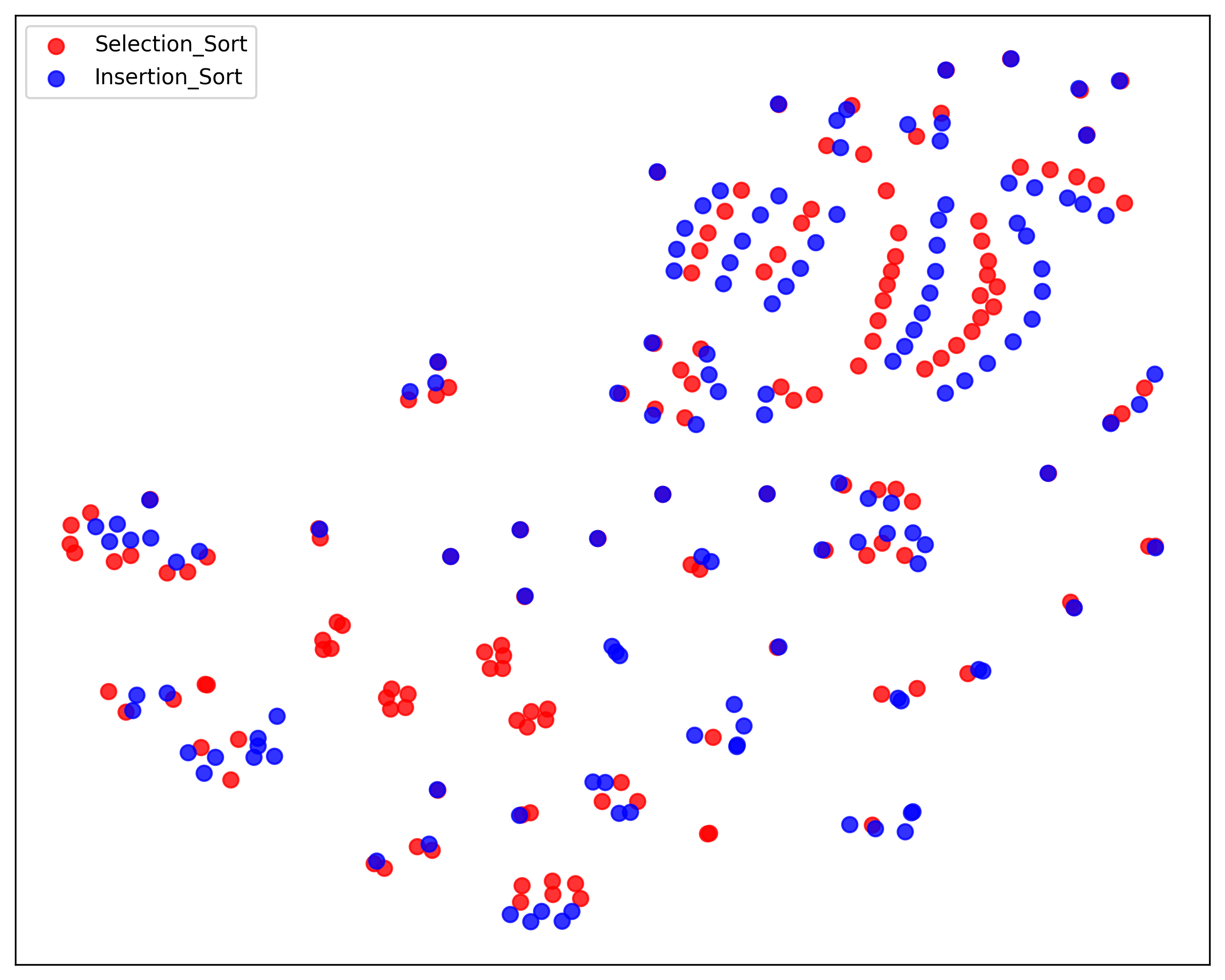}
        \caption{Selection Sort vs. Insertion Sort}
    \end{subfigure}

    \caption{Pairwise comparisons of classical sorting algorithms using t-SNE, showing the token embeddings in a 2D space (Part 1).}
    \label{fig:sorting-algorithms-comparisons-tsne-part1}
\end{figure}

\begin{figure}[htbp]
    \centering

    \begin{subfigure}{0.48\textwidth}
        \includegraphics[width=\textwidth]{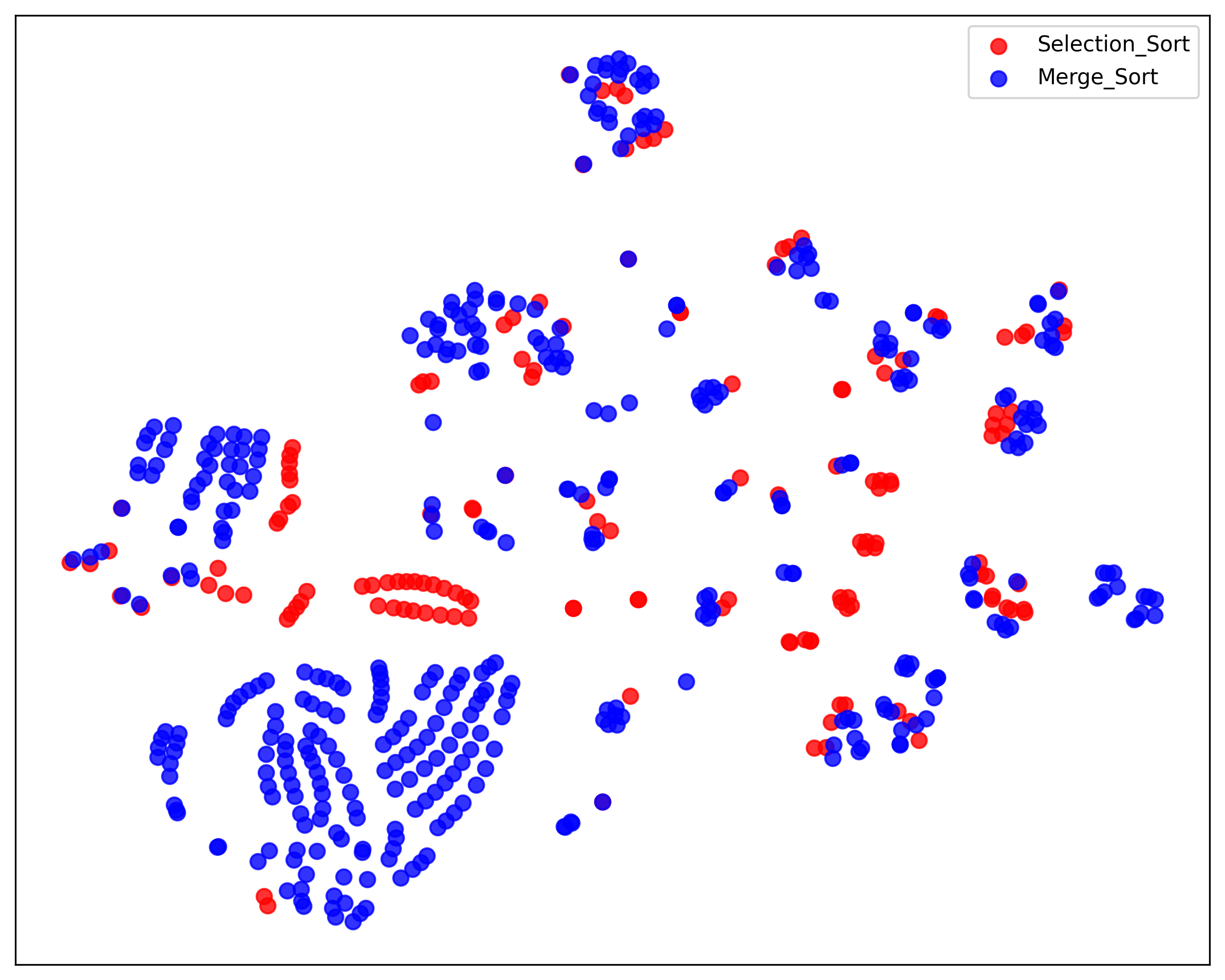}
        \caption{Selection Sort vs. Merge Sort}
    \end{subfigure}
    \hfill
    \begin{subfigure}{0.48\textwidth}
        \includegraphics[width=\textwidth]{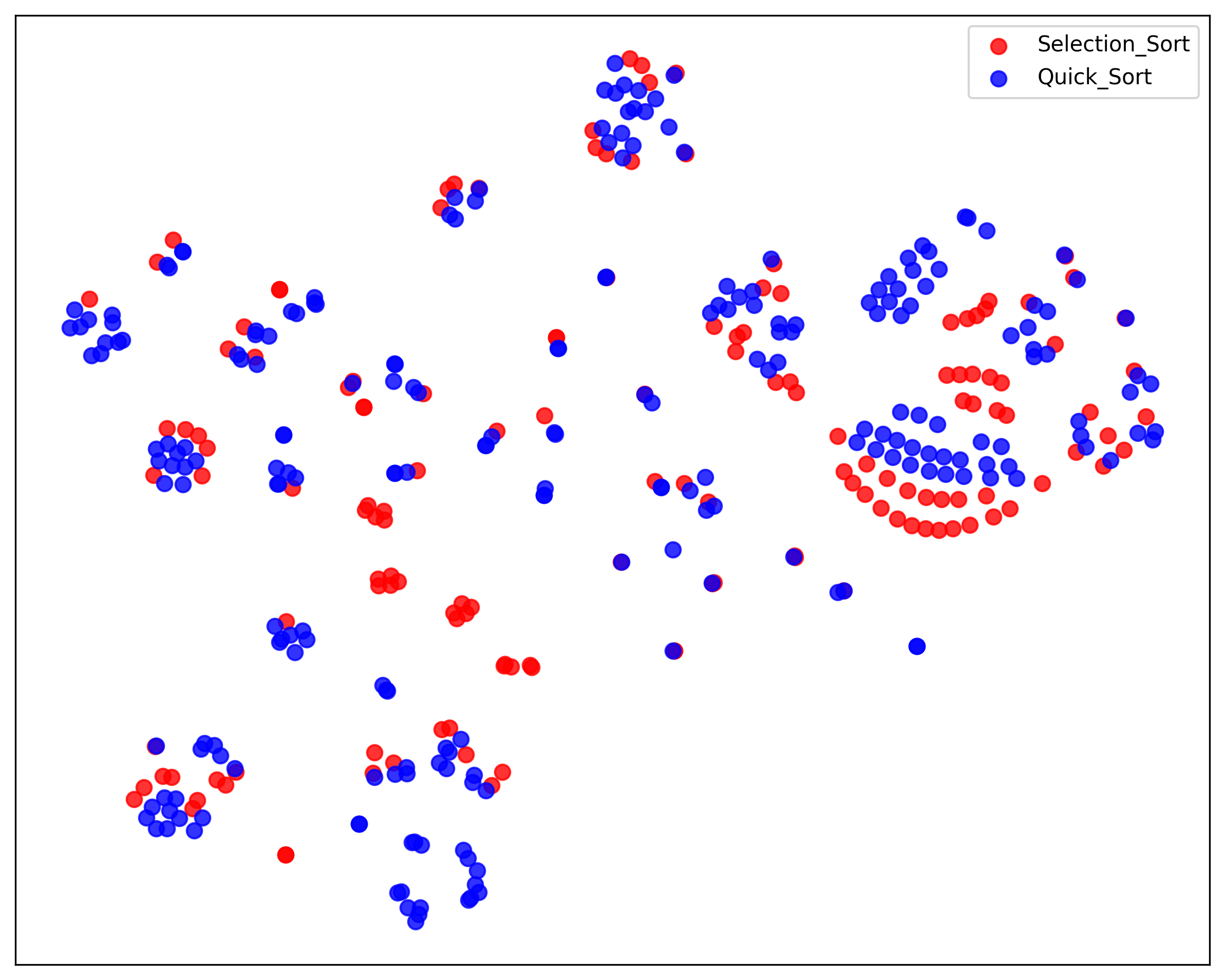}
        \caption{Selection Sort vs. Quick Sort}
    \end{subfigure}

    \vspace{0.3cm}

    \begin{subfigure}{0.48\textwidth}
        \includegraphics[width=\textwidth]{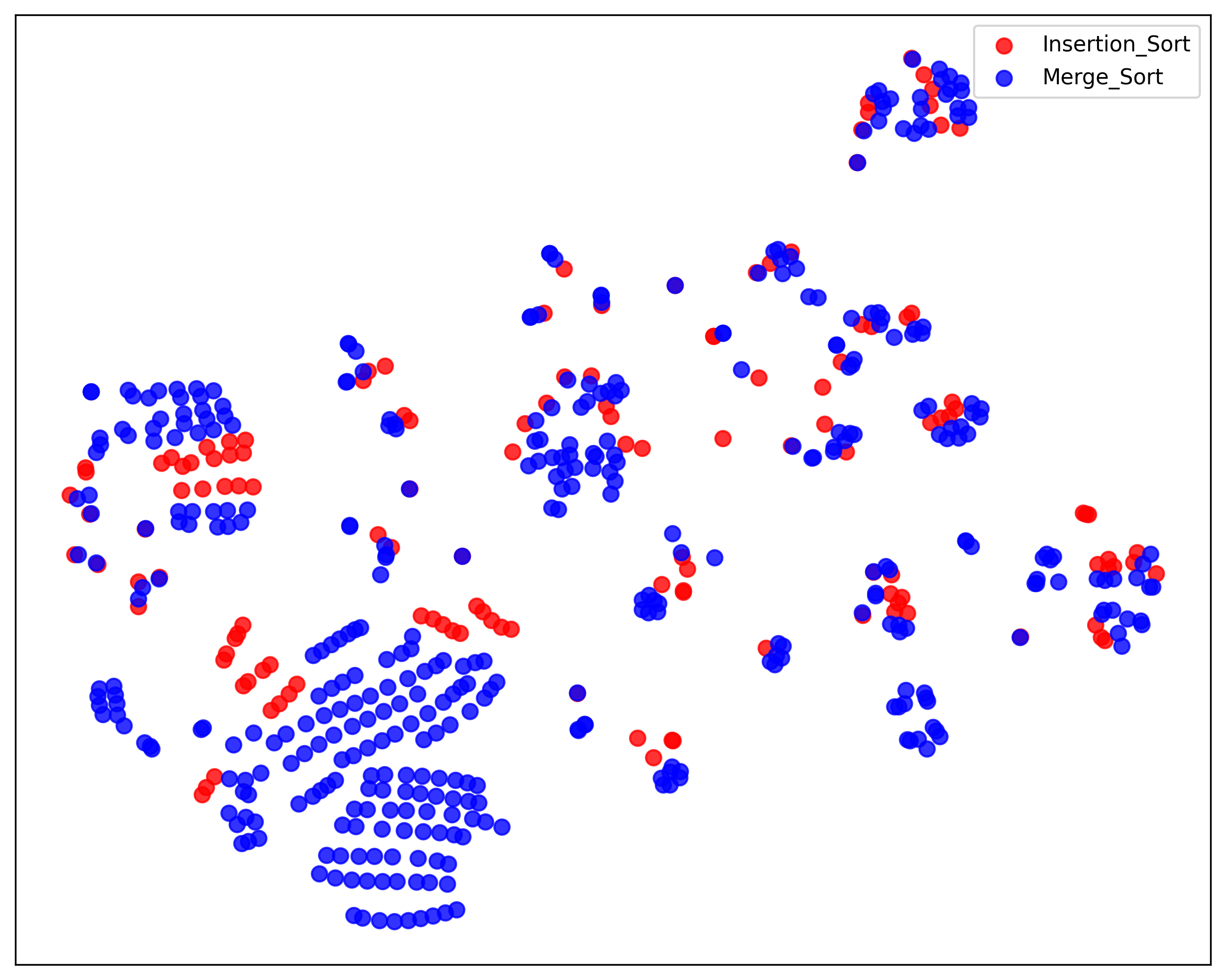}
        \caption{Insertion Sort vs. Merge Sort}
    \end{subfigure}
    \hfill
    \begin{subfigure}{0.48\textwidth}
        \includegraphics[width=\textwidth]{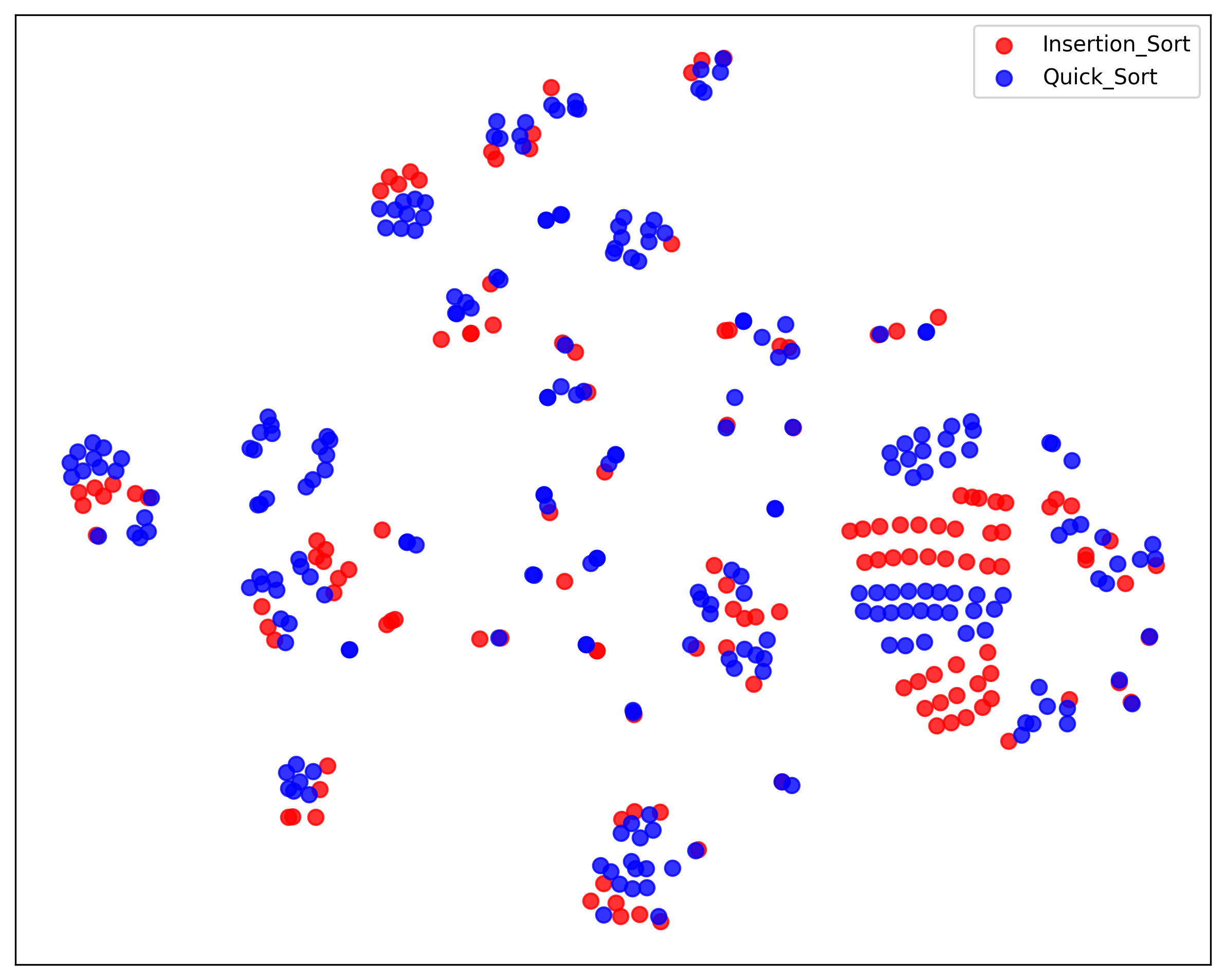}
        \caption{Insertion Sort vs. Quick Sort}
    \end{subfigure}

    \vspace{0.3cm}

    \begin{subfigure}{0.48\textwidth}
        \includegraphics[width=\textwidth]{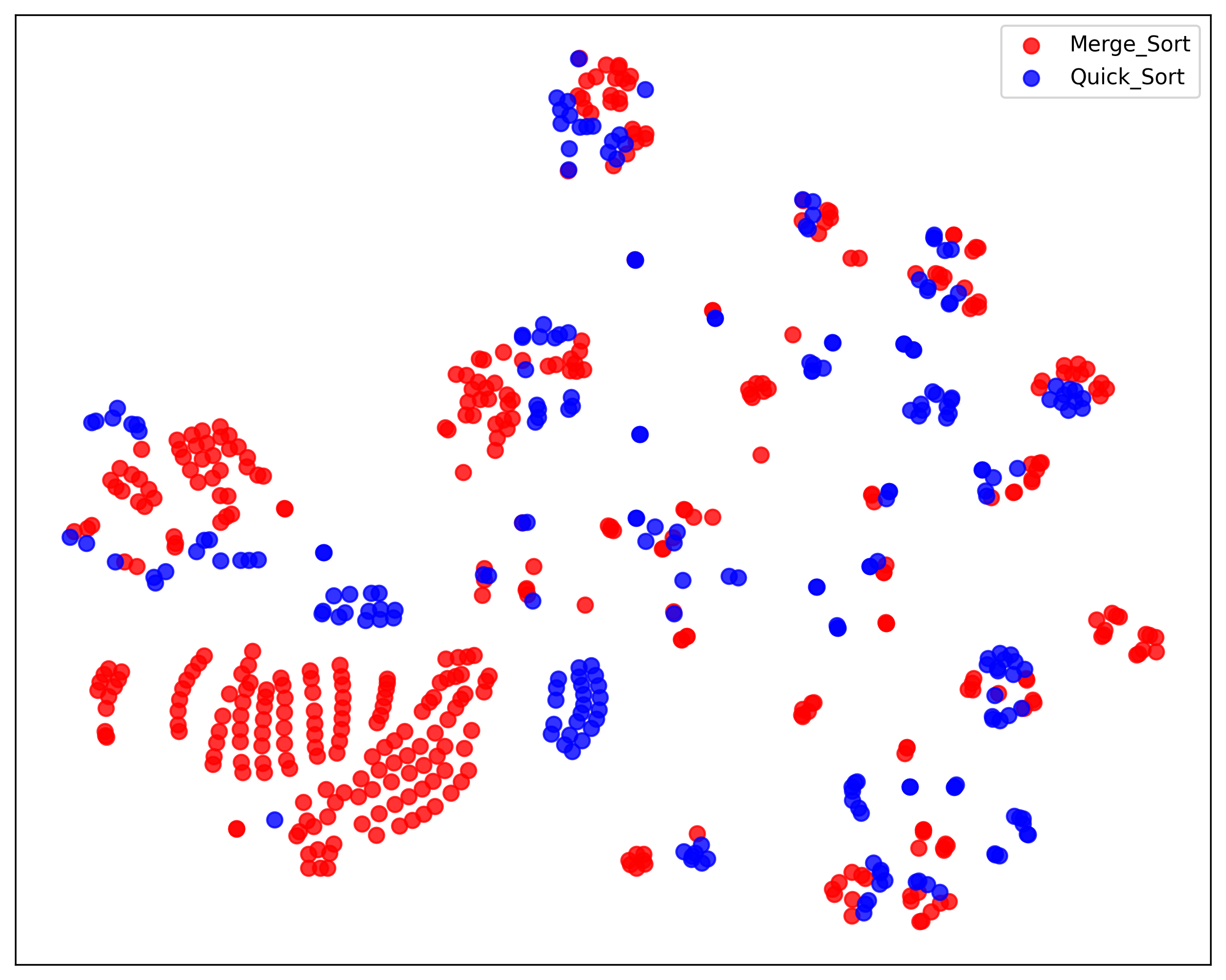}
        \caption{Merge Sort vs. Quick Sort}
    \end{subfigure}

    \caption{Pairwise comparisons of classical sorting algorithms using t-SNE, showing the token embeddings in a 2D space (Part 2).}
    \label{fig:sorting-algorithms-comparisons-tsne-part2}
\end{figure}

Figures \ref{fig:sorting-algorithms-comparisons-tsne-part1} and \ref{fig:sorting-algorithms-comparisons-tsne-part2} illustrate these pairwise comparisons, helping to reveal potential similarities or differences in the way the algorithms are represented through their token embeddings.

\subsection{Pairwise Comparisons of Sorting Algorithms using UMAP}
We also present pairwise comparisons between classical sorting algorithms by visualizing their token-level embeddings in a two-dimensional space using UMAP. This method allows for preserving the global and local structures of the high-dimensional token embeddings when projected in 2D. Once again, each image shows the distribution of token embeddings for a specific pair of sorting algorithms, with different colors distinguishing the algorithms. These visualizations provide a deeper look into how the token embeddings for each algorithm are organized and how their representations relate to one another.

\begin{figure}[htbp]
    \centering

    \begin{subfigure}{0.48\textwidth}
        \includegraphics[width=\textwidth]{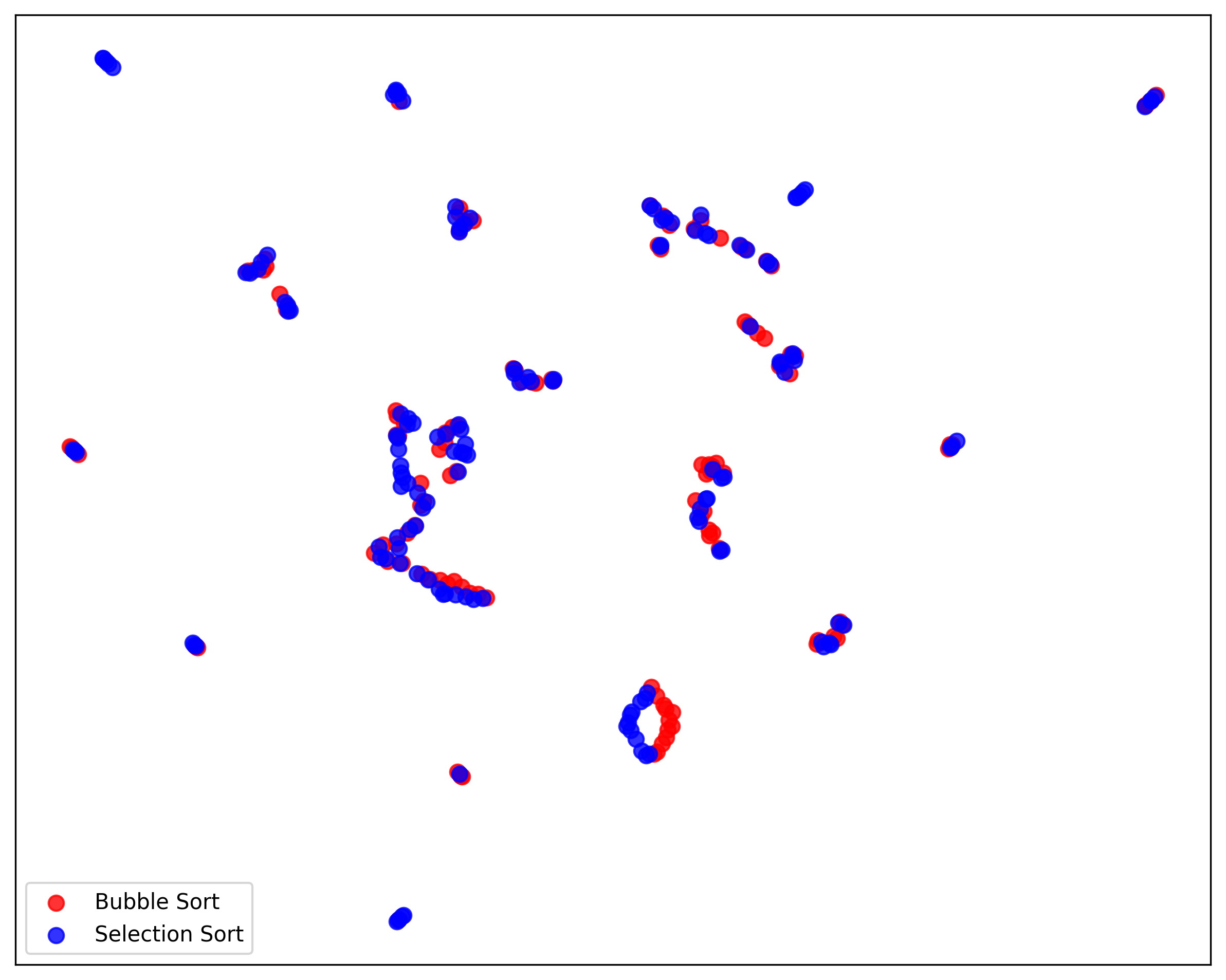}
        \caption{Bubble Sort vs. Selection Sort}
    \end{subfigure}
    \hfill
    \begin{subfigure}{0.48\textwidth}
        \includegraphics[width=\textwidth]{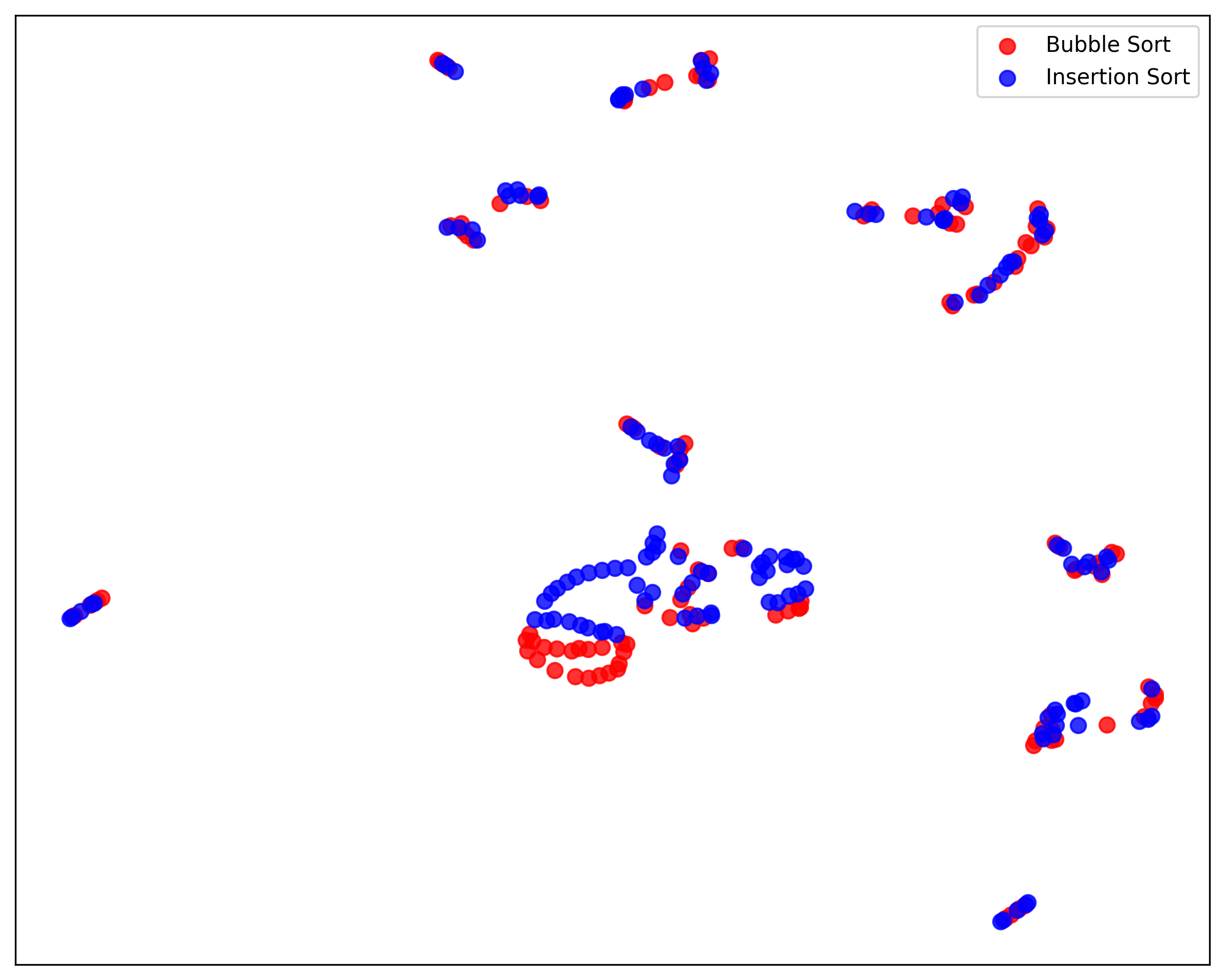}
        \caption{Bubble Sort vs. Insertion Sort}
    \end{subfigure}

    \vspace{0.3cm}

    \begin{subfigure}{0.48\textwidth}
        \includegraphics[width=\textwidth]{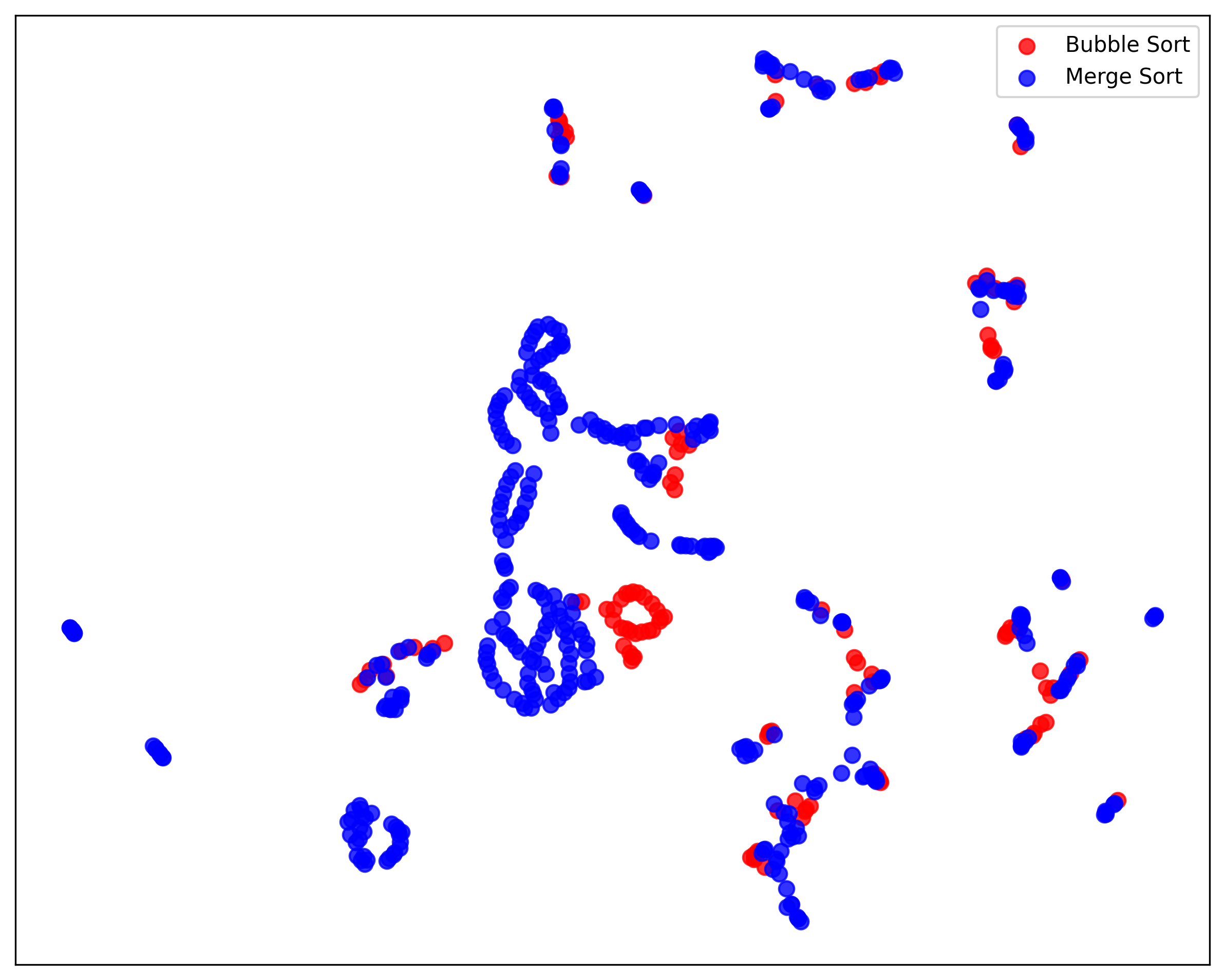}
        \caption{Bubble Sort vs. Merge Sort}
    \end{subfigure}
    \hfill
    \begin{subfigure}{0.48\textwidth}
        \includegraphics[width=\textwidth]{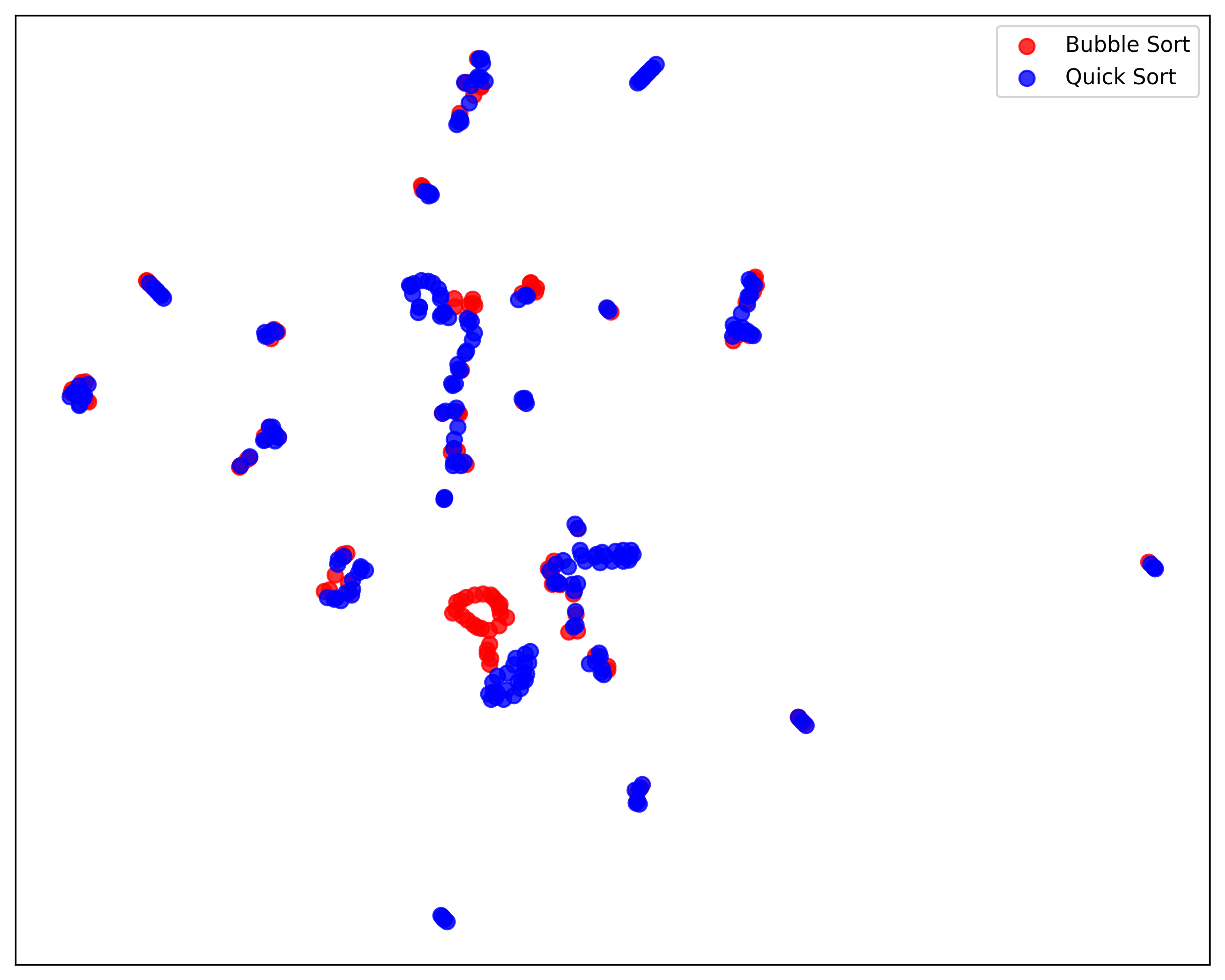}
        \caption{Bubble Sort vs. Quick Sort}
    \end{subfigure}

    \vspace{0.3cm}

    \begin{subfigure}{0.48\textwidth}
        \includegraphics[width=\textwidth]{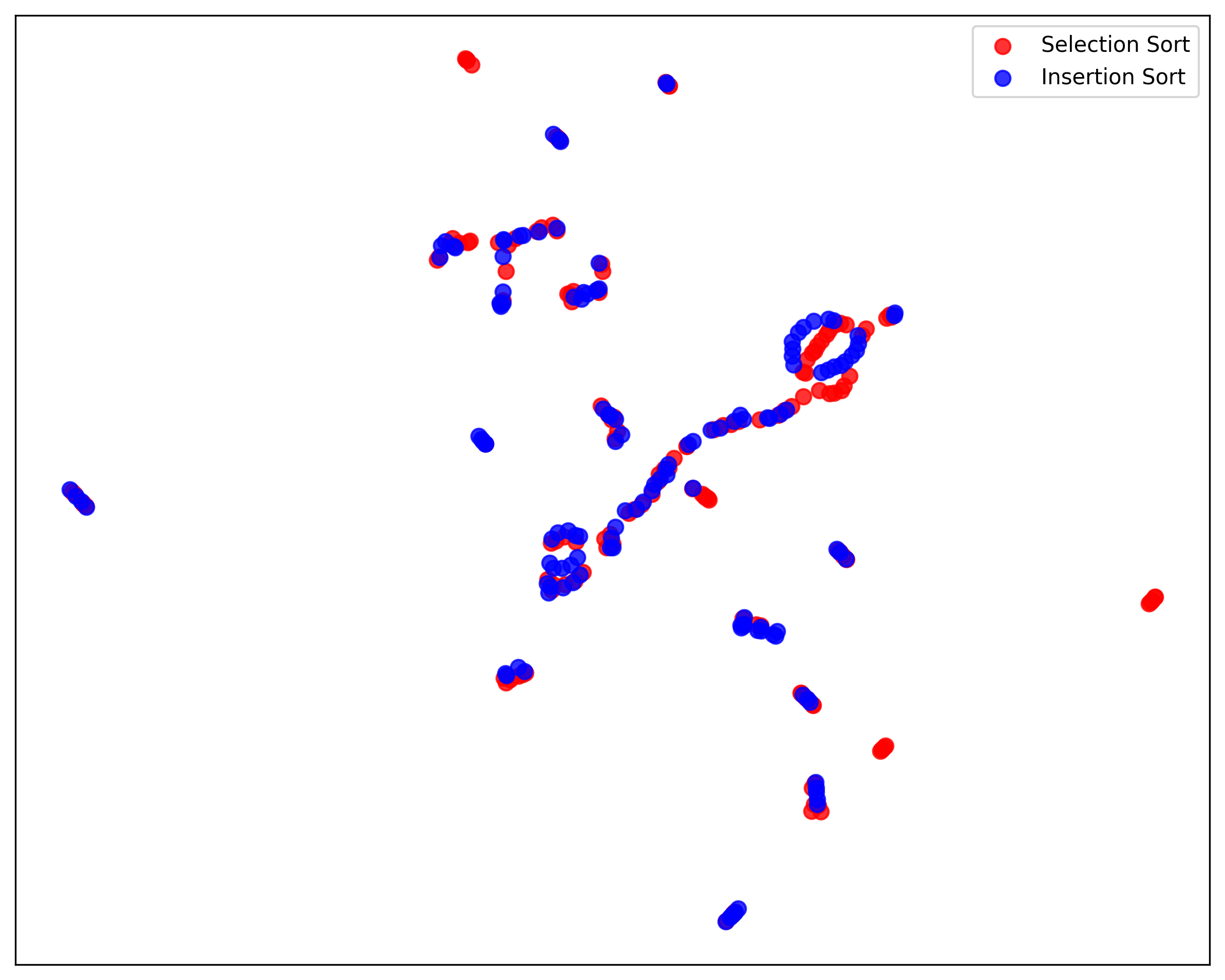}
        \caption{Selection Sort vs. Insertion Sort}
    \end{subfigure}

    \caption{Pairwise comparisons of classical sorting algorithms using UMAP, showing the token-level embeddings in a 2D space (Part 1).}
    \label{fig:umap-comparisons-part1}
\end{figure}

\begin{figure}[htbp]
    \centering

    \begin{subfigure}{0.48\textwidth}
        \includegraphics[width=\textwidth]{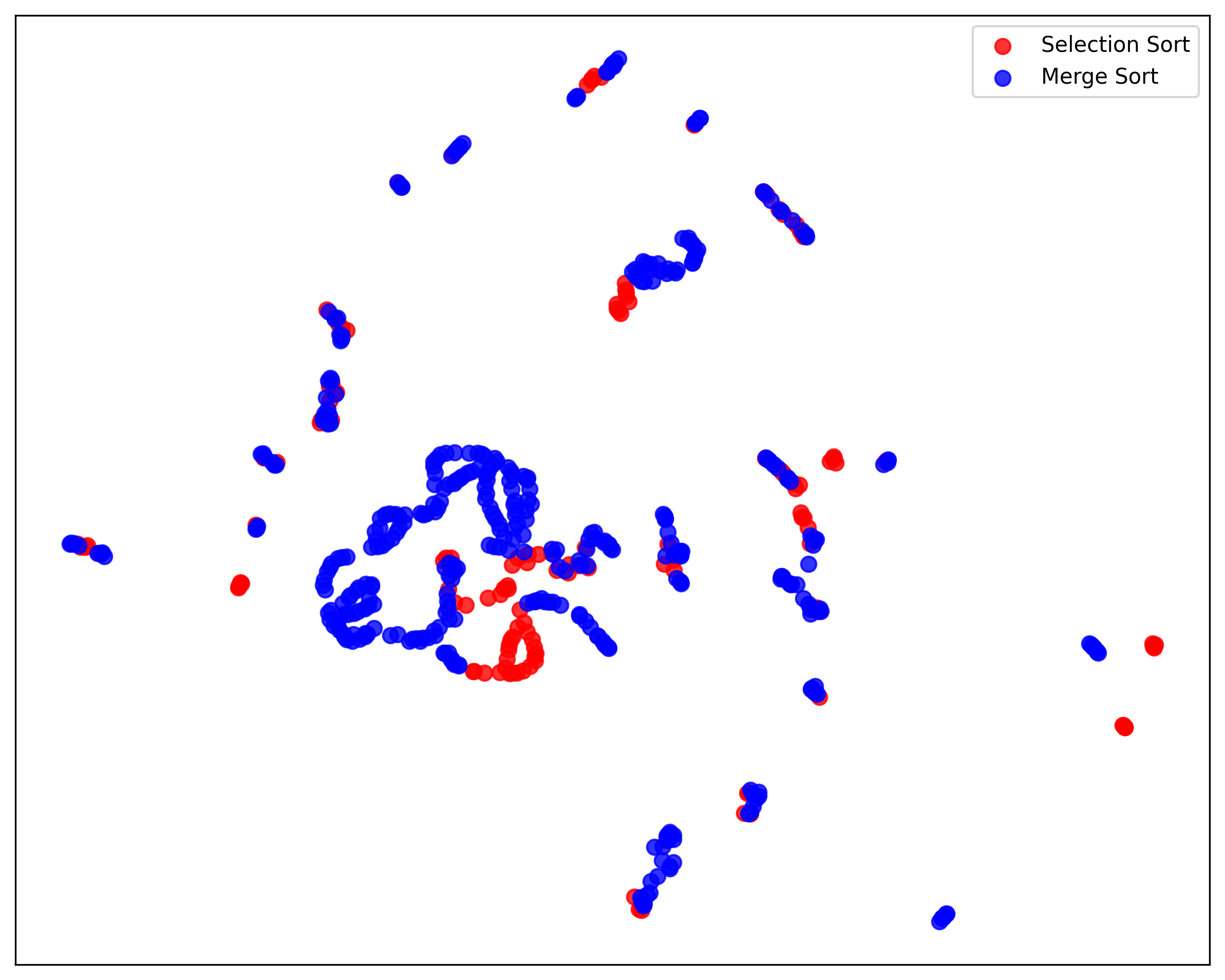}
        \caption{Selection Sort vs. Merge Sort}
    \end{subfigure}
    \hfill
    \begin{subfigure}{0.48\textwidth}
        \includegraphics[width=\textwidth]{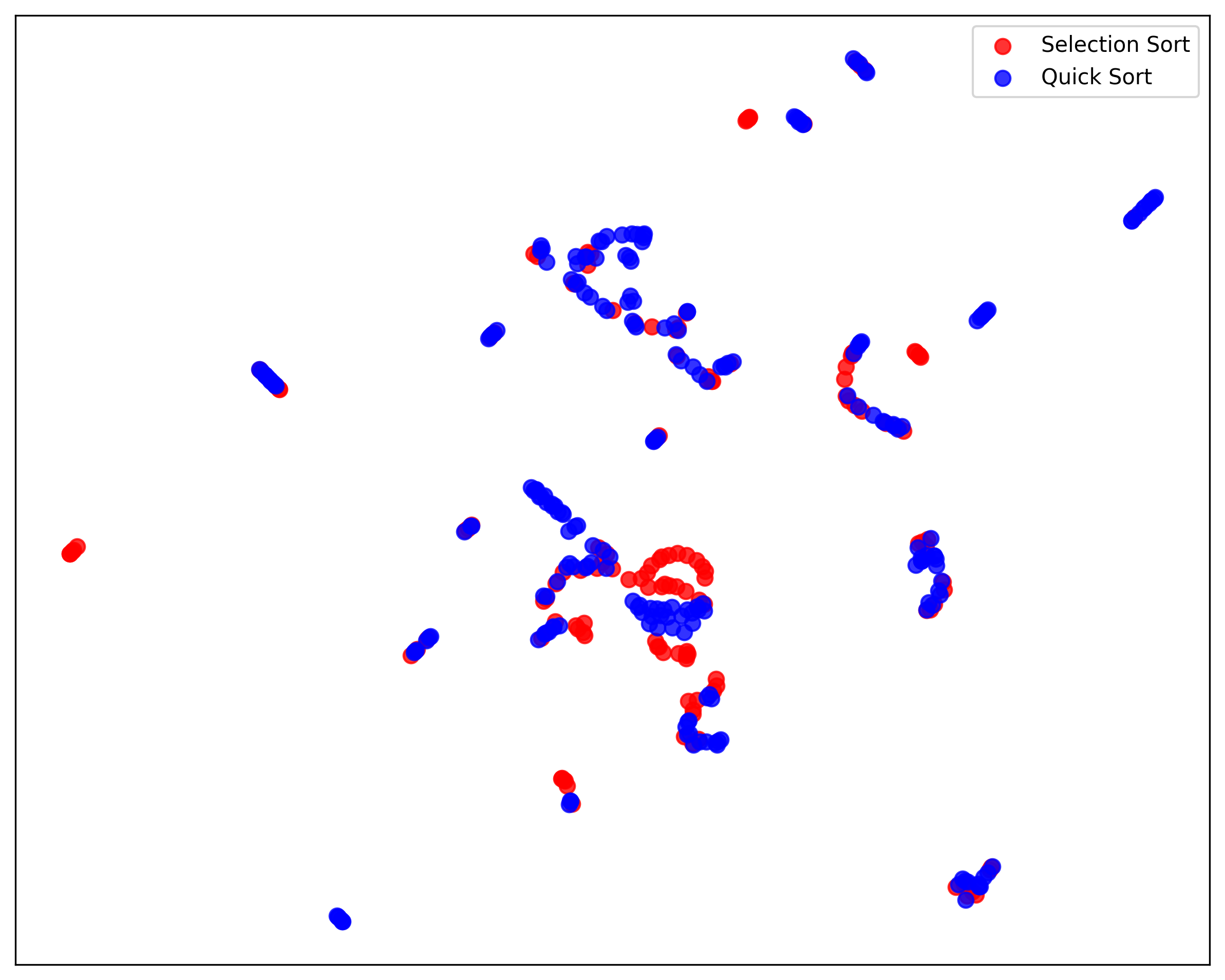}
        \caption{Selection Sort vs. Quick Sort}
    \end{subfigure}

    \vspace{0.3cm}

    \begin{subfigure}{0.48\textwidth}
        \includegraphics[width=\textwidth]{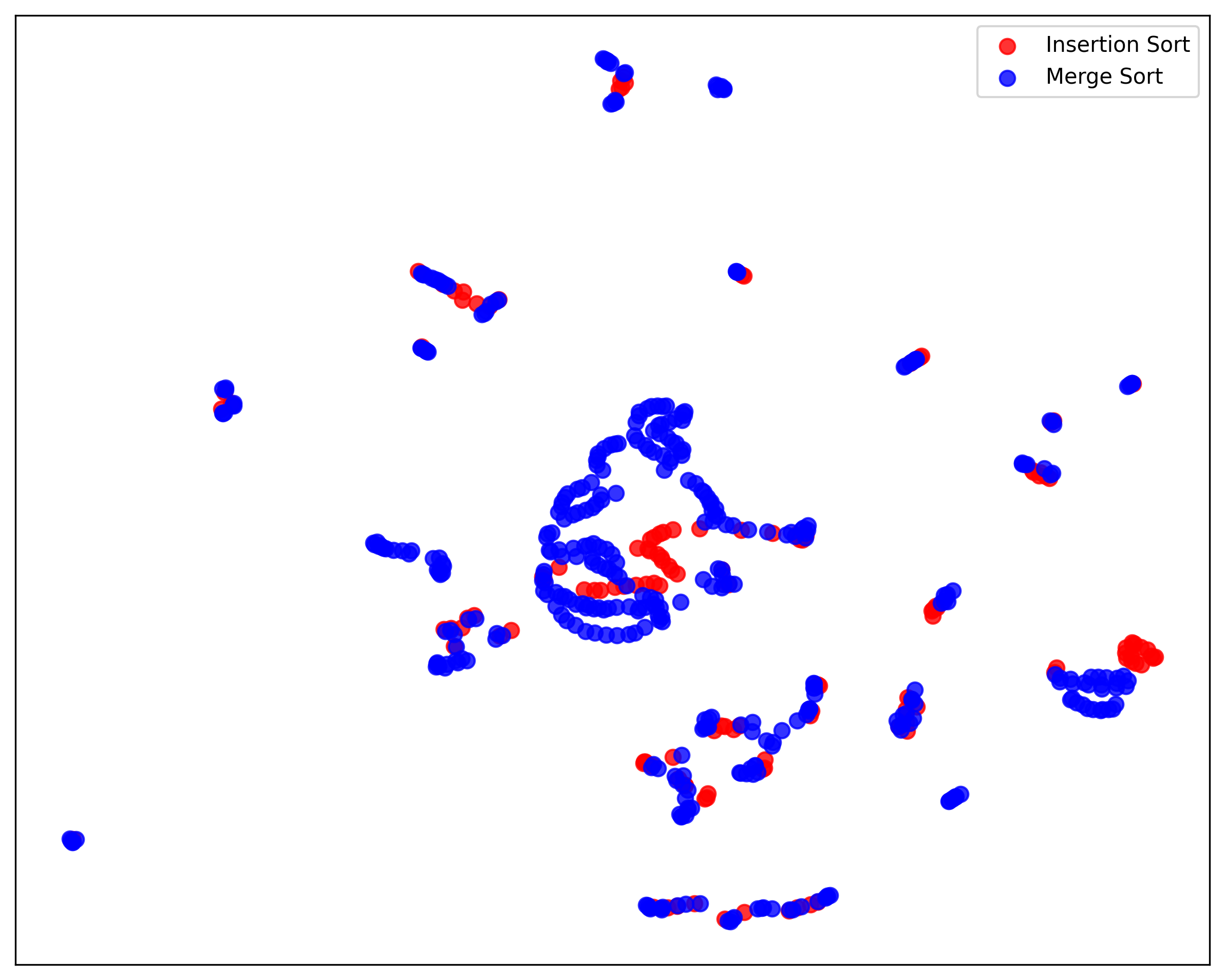}
        \caption{Insertion Sort vs. Merge Sort}
    \end{subfigure}
    \hfill
    \begin{subfigure}{0.48\textwidth}
        \includegraphics[width=\textwidth]{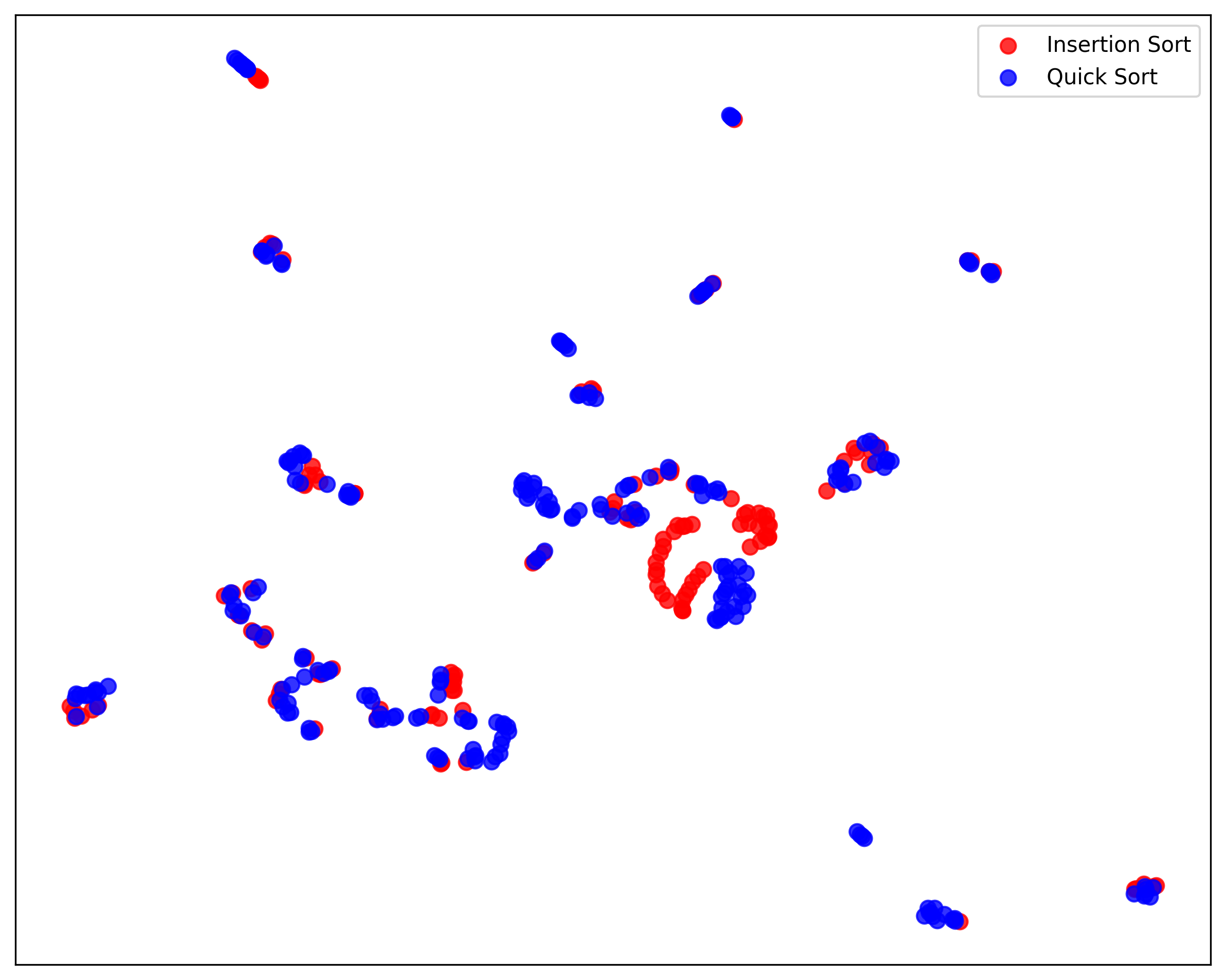}
        \caption{Insertion Sort vs. Quick Sort}
    \end{subfigure}

    \vspace{0.3cm}

    \begin{subfigure}{0.48\textwidth}
        \includegraphics[width=\textwidth]{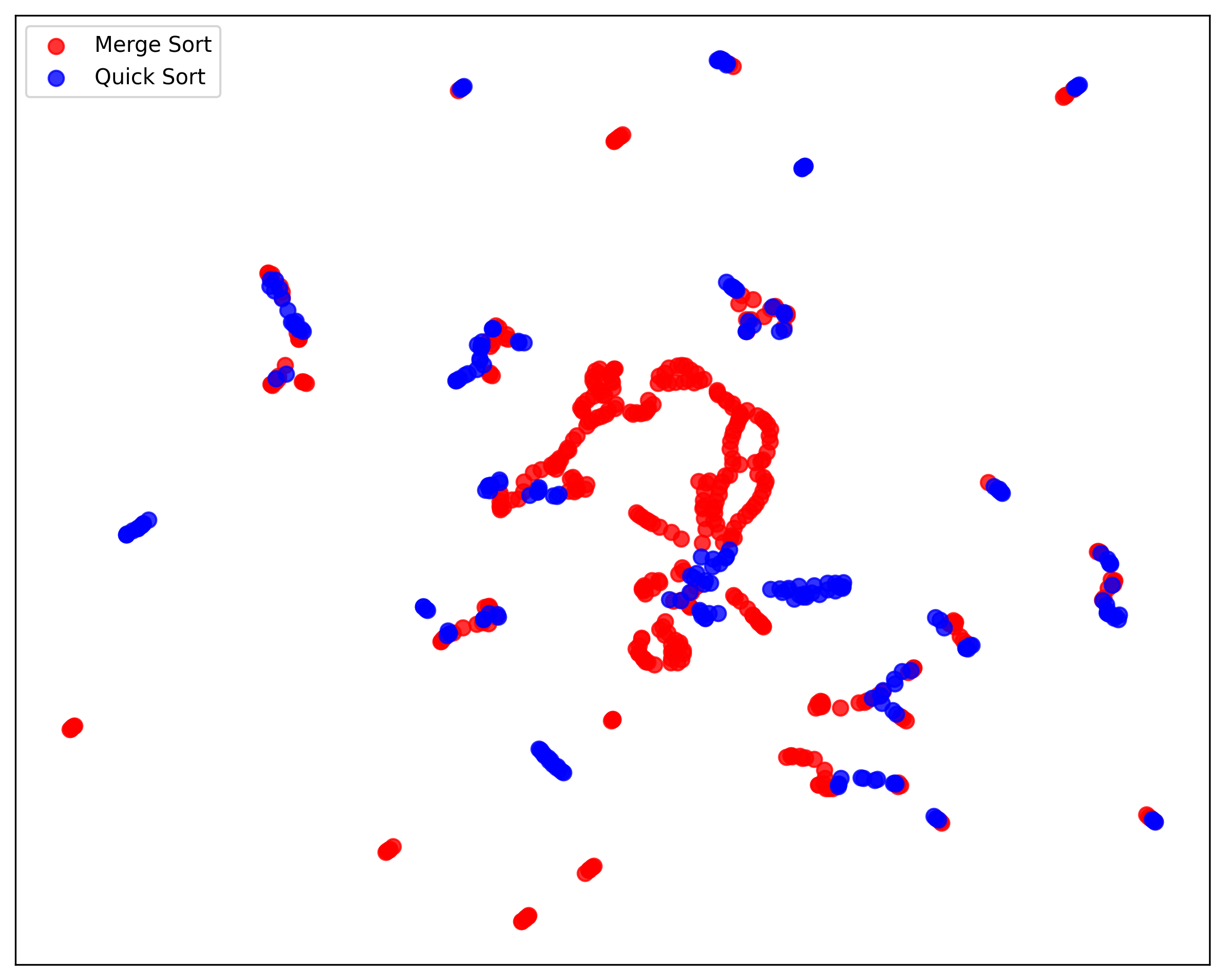}
        \caption{Merge Sort vs. Quick Sort}
    \end{subfigure}

    \caption{Pairwise comparisons of classical sorting algorithms using UMAP, showing the token-level embeddings in a 2D space (Part 2).}
    \label{fig:umap-comparisons-part2}
\end{figure}

Figures \ref{fig:umap-comparisons-part1} and \ref{fig:umap-comparisons-part2} display these pairwise comparisons, shedding light on potential structural patterns or distinctions among the token embeddings of the algorithms.

\subsection{Saliency Maps for Sorting Algorithms}
\begin{figure}[H]
    \centering
    \begin{subfigure}[b]{0.48\textwidth}
        \centering
        \includegraphics[width=\textwidth]{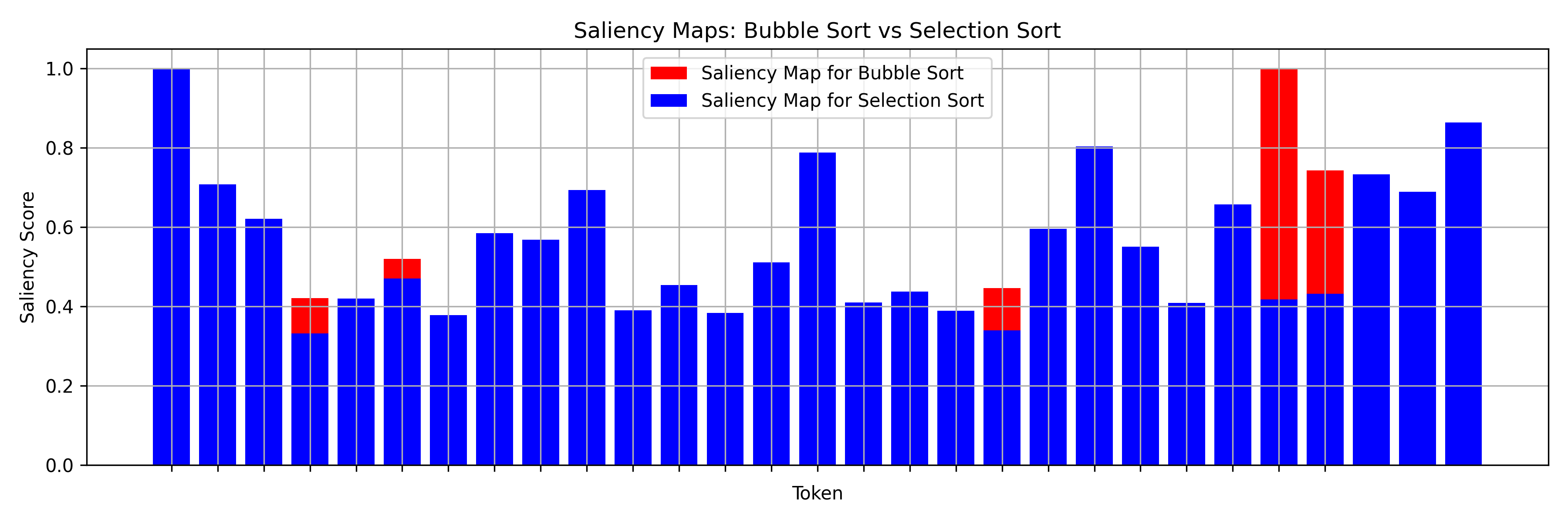}
        \caption{Bubble Sort vs Selection Sort}
        \label{fig:bubble_vs_selection}
    \end{subfigure}
    \hfill
    \begin{subfigure}[b]{0.48\textwidth}
        \centering
        \includegraphics[width=\textwidth]{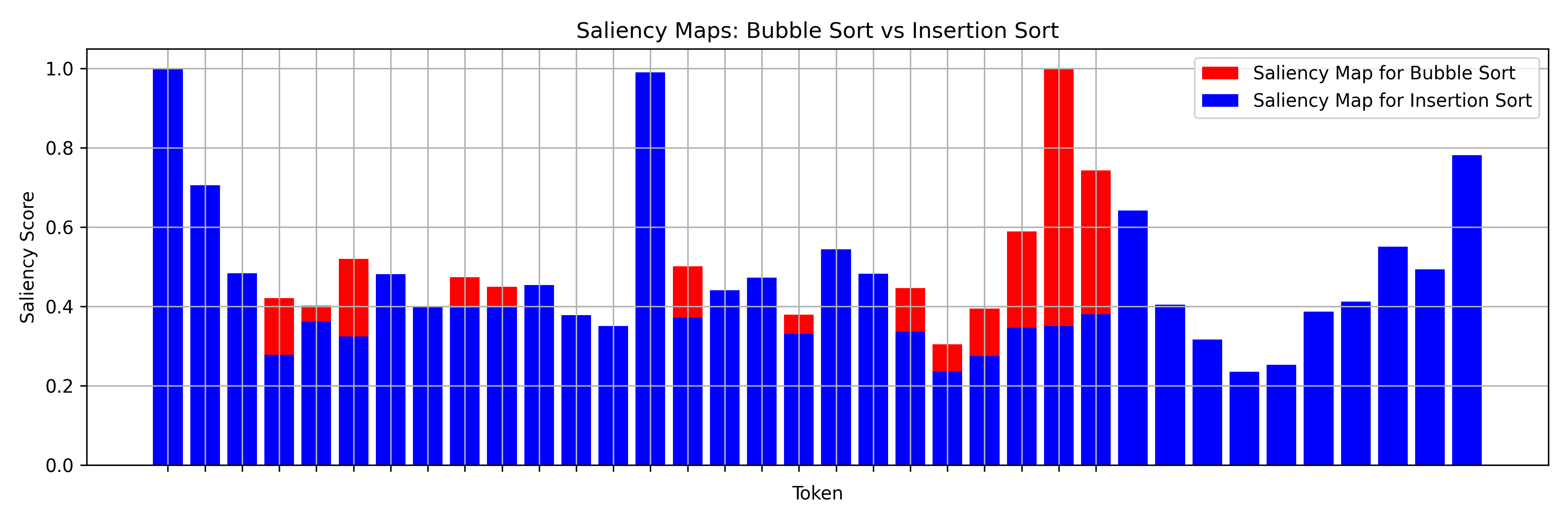}
        \caption{Bubble Sort vs Insertion Sort}
        \label{fig:bubble_vs_insertion}
    \end{subfigure}
    
    \vspace{0.5cm}
    
    \begin{subfigure}[b]{0.48\textwidth}
        \centering
        \includegraphics[width=\textwidth]{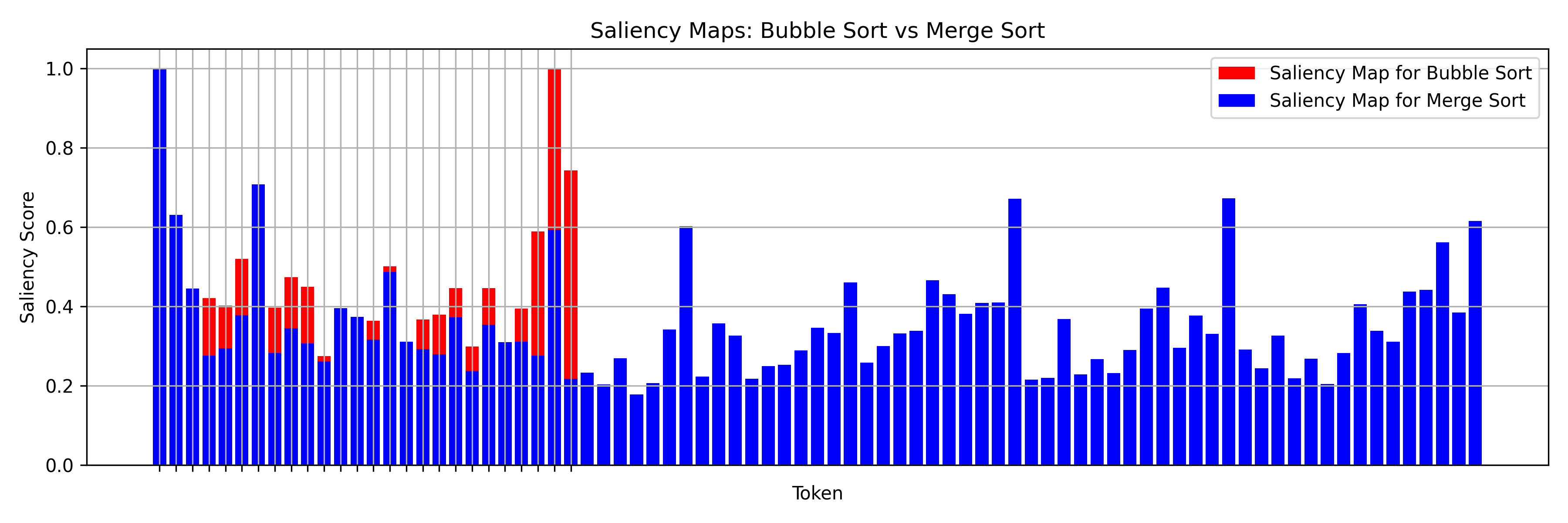}
        \caption{Bubble Sort vs Merge Sort}
        \label{fig:bubble_vs_merge}
    \end{subfigure}
    \hfill
    \begin{subfigure}[b]{0.48\textwidth}
        \centering
        \includegraphics[width=\textwidth]{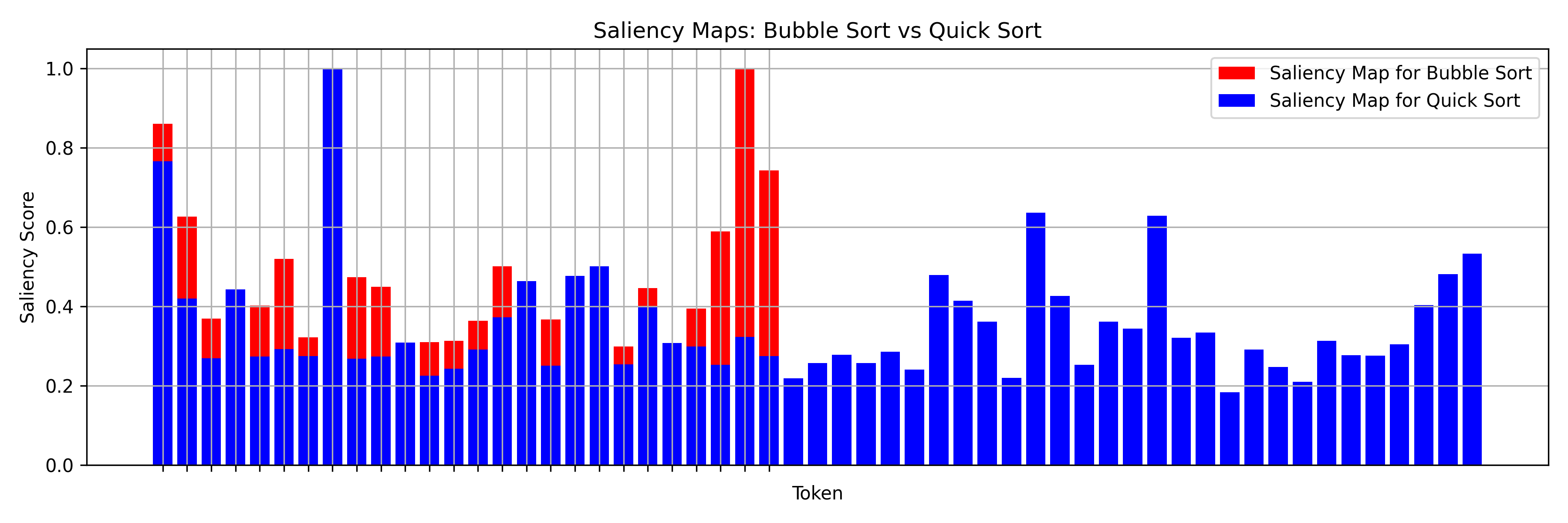}
        \caption{Bubble Sort vs Quick Sort}
        \label{fig:bubble_vs_quick}
    \end{subfigure}
    
    \vspace{0.5cm}
    
    \begin{subfigure}[b]{0.48\textwidth}
        \centering
        \includegraphics[width=\textwidth]{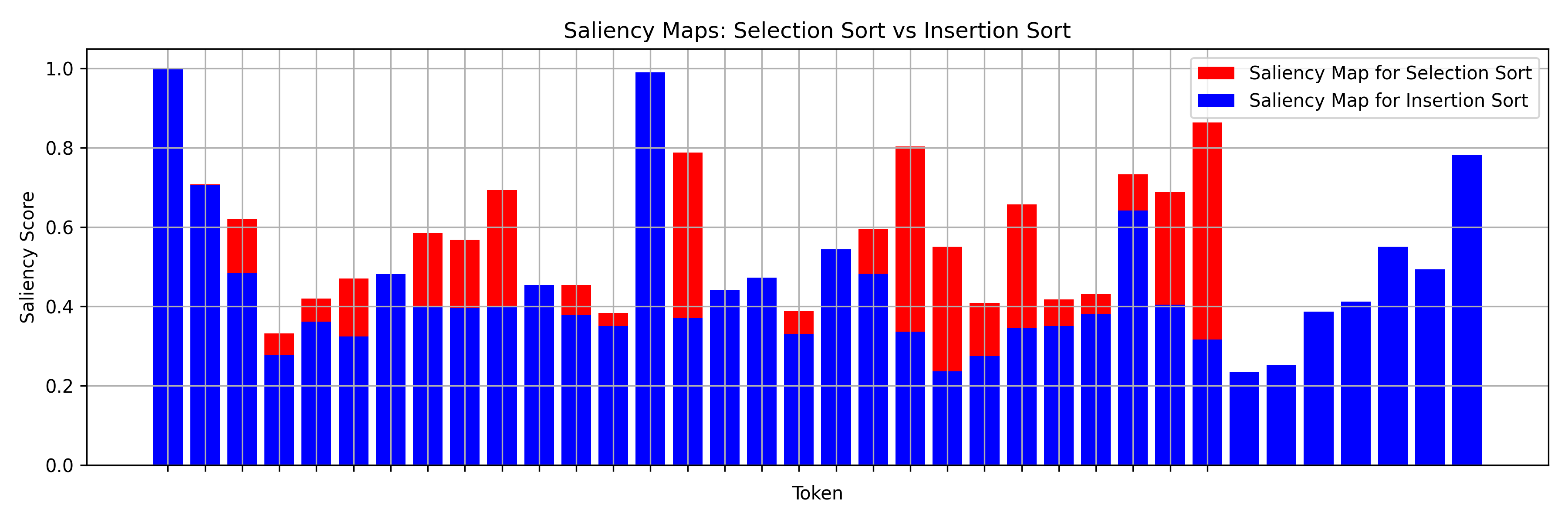}
        \caption{Selection Sort vs Insertion Sort}
        \label{fig:selection_vs_insertion}
    \end{subfigure}
    \hfill
    \begin{subfigure}[b]{0.48\textwidth}
        \centering
        \includegraphics[width=\textwidth]{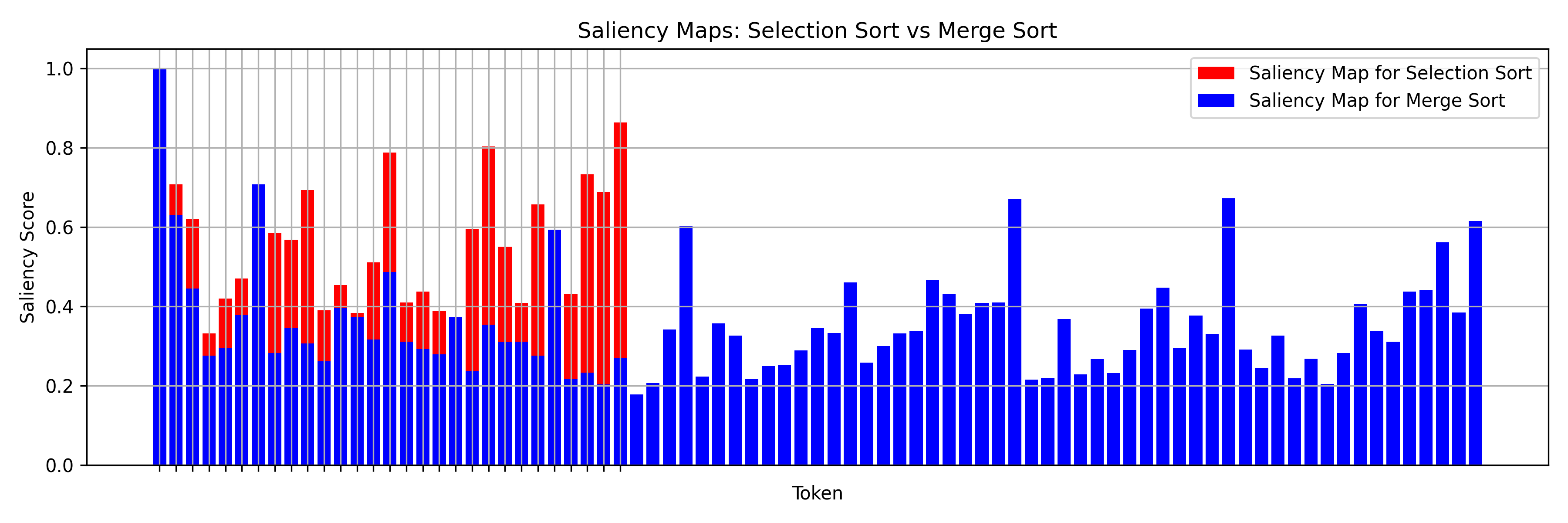}
        \caption{Selection Sort vs Merge Sort}
        \label{fig:selection_vs_merge}
    \end{subfigure}
    
    \vspace{0.5cm}
    
    \begin{subfigure}[b]{0.48\textwidth}
        \centering
        \includegraphics[width=\textwidth]{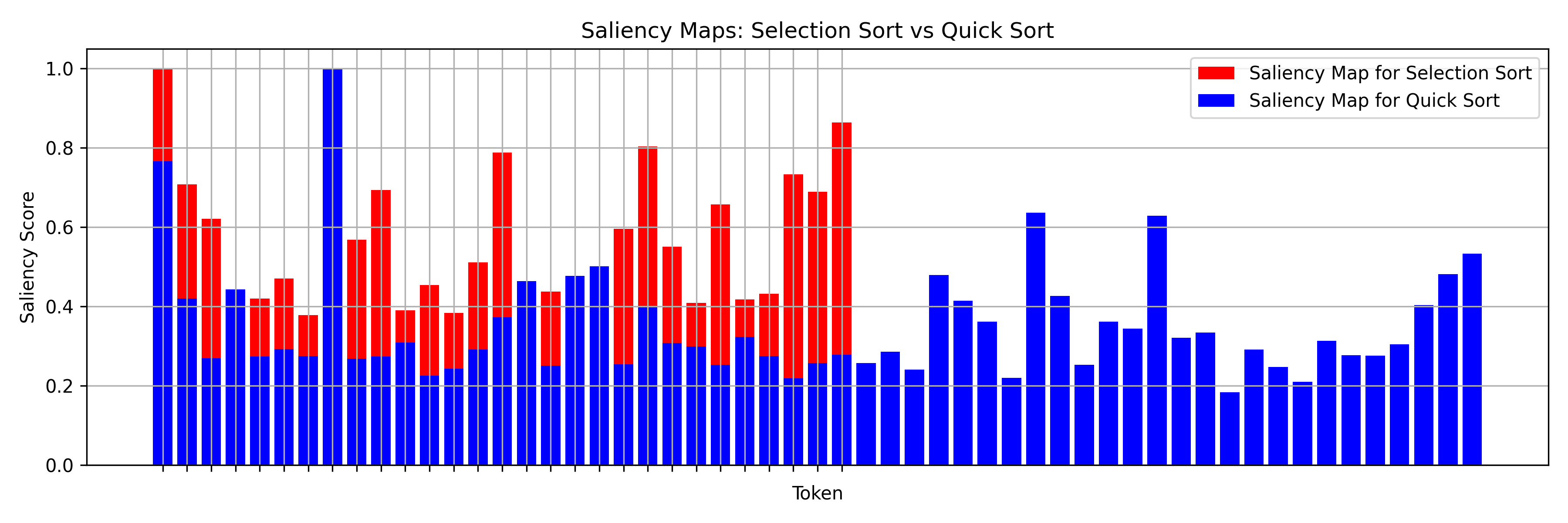}
        \caption{Selection Sort vs Quick Sort}
        \label{fig:selection_vs_quick}
    \end{subfigure}
    \hfill
    \begin{subfigure}[b]{0.48\textwidth}
        \centering
        \includegraphics[width=\textwidth]{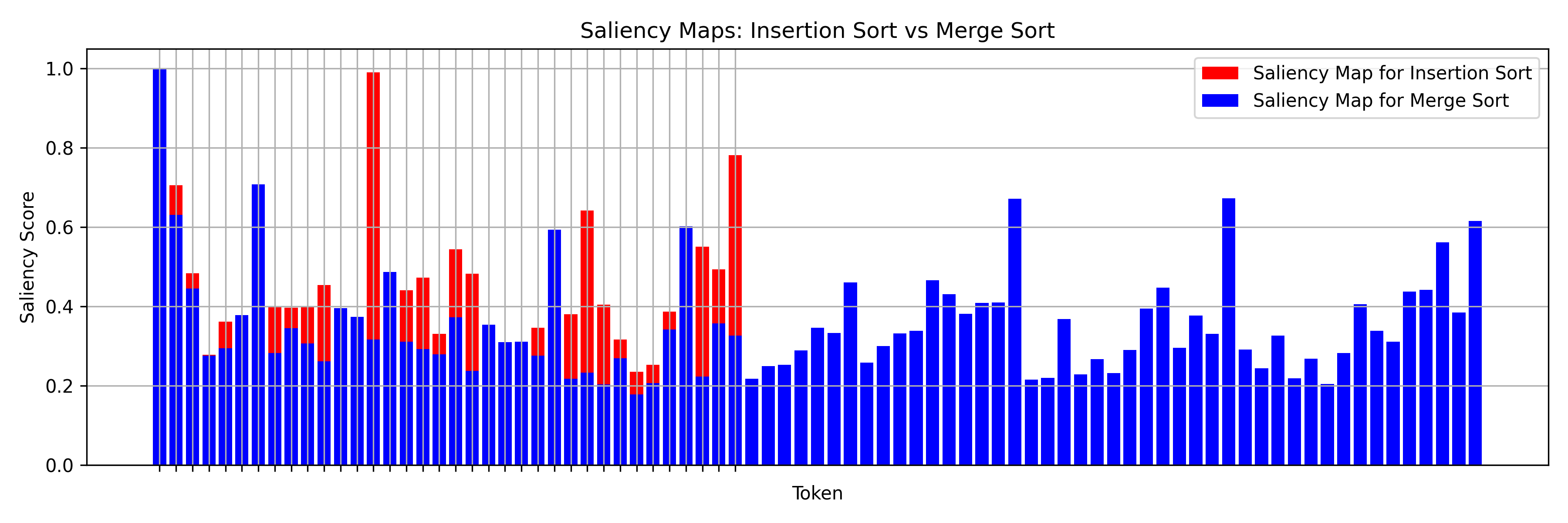}
        \caption{Insertion Sort vs Merge Sort}
        \label{fig:insertion_vs_merge}
    \end{subfigure}
    
    \vspace{0.5cm}
    
    \begin{subfigure}[b]{0.48\textwidth}
        \centering
        \includegraphics[width=\textwidth]{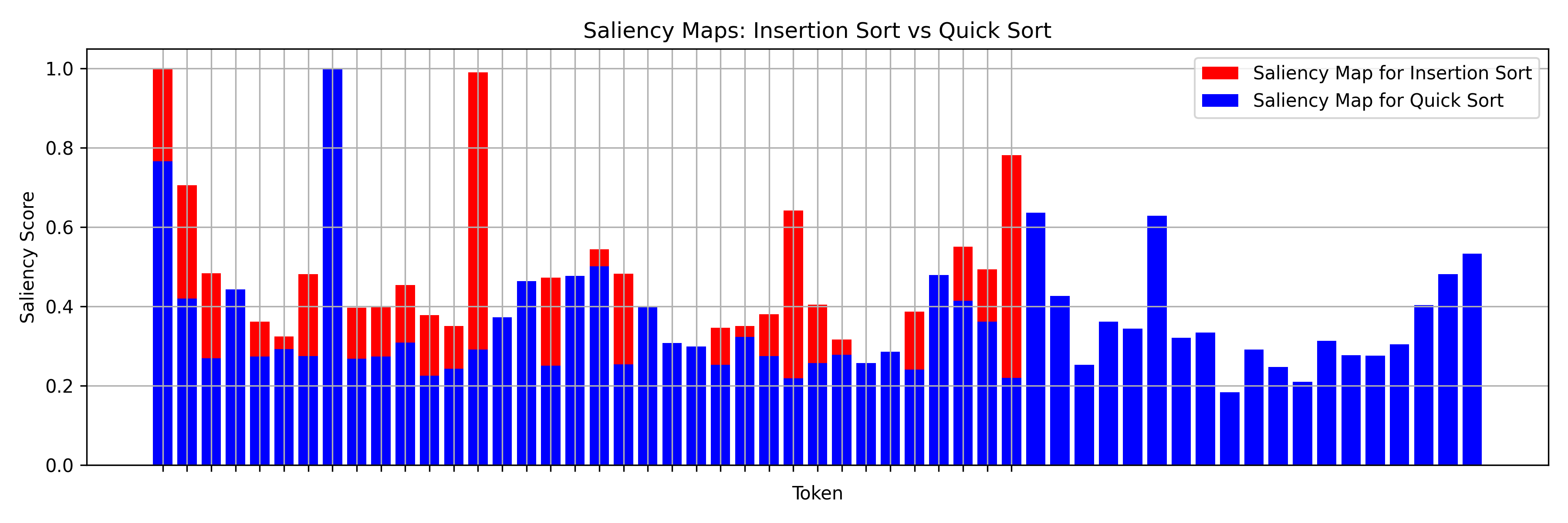}
        \caption{Insertion Sort vs Quick Sort}
        \label{fig:insertion_vs_quick}
    \end{subfigure}
    \hfill
    \begin{subfigure}[b]{0.48\textwidth}
        \centering
        \includegraphics[width=\textwidth]{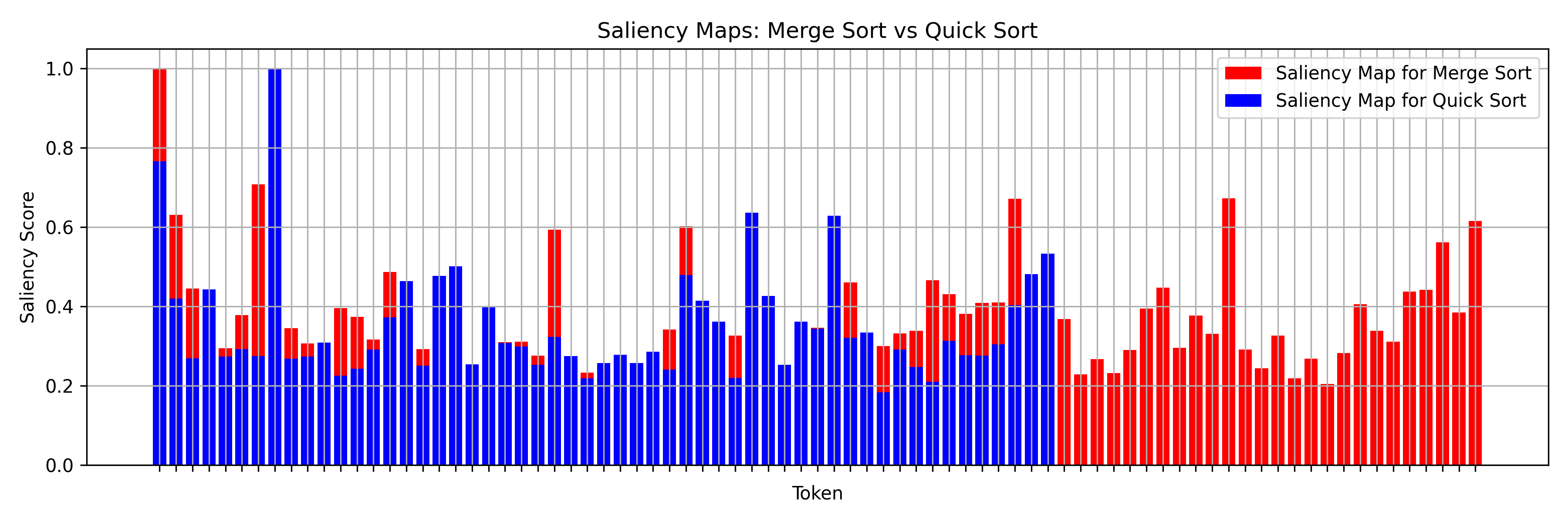}
        \caption{Merge Sort vs Quick Sort}
        \label{fig:merge_vs_quick}
    \end{subfigure}
    
    \caption{Saliency maps for pairwise comparisons of different sorting algorithms.}
    \label{fig:saliency_maps}
\end{figure}

We also present the saliency maps generated for the pairwise comparisons of different sorting algorithms using the GraphCodeBERT model. Saliency maps provide a visual explanation of which parts of the input contributed most to the model's decisions, offering hints into how the model interprets the similarities between the algorithms. These maps allow us to observe which tokens or features the model focuses on when distinguishing between the sorting algorithms. Figure \ref{fig:saliency_maps} illustrates the saliency maps for these comparisons, focusing on the critical areas of attention within the token embeddings.

The results show that this method holds significant potential for various applications. For example, automated code reviews help reviewers quickly identify redundancies, ensuring consistency and reducing maintenance costs. In refactoring, this method can assist developers in identifying code segments that can be optimized. Additionally, in educational settings, this approach can aid in teaching by offering a deeper understanding of the underlying principles of algorithms. Furthermore, the ability to detect code similarities has promising applications in plagiarism detection, where it can identify functionally equivalent code that has been superficially altered to evade detection.

\section{Discussion}
Our strategy can improve interpretability by offering different approaches to visualize semantic similarities between code fragments through the attention mechanism of GraphCodeBERT. We have seen that this approach improves over traditional syntactic comparison approaches, providing deeper insights into the functional and logical similarities between code fragments.

However, several areas warrant further investigation. One key area is the method's scalability when applied to larger codebases. While our initial experiments focused on relatively small examples, real-world software projects involve thousands or even millions of lines of code. Assessing how well the model performs under such conditions, both in terms of computational efficiency and the quality of the generated outputs, will be crucial.

Another important aspect is the method's applicability across different programming languages. While our experiments showed promising results in single-language comparisons, expanding the range of languages and including those with other structures (e.g., functional versus object-oriented programming) would provide a more comprehensive evaluation of the method's usefulness.

Future work may also explore integrating this approach with other code analysis methods. For instance, combining our method with static analysis or code metrics tools could provide a more holistic view of code maintainability. Additionally, incorporating feedback mechanisms where developers can annotate or refine the outputs could improve the accuracy and usability of the results.

\section{Conclusion}
Our research introduces a novel strategy for increasing the interpretability capabilities of the similarity between code fragments using GraphCodeBERT. The process is formalized through mathematical expressions and attention mechanisms, presenting a framework that captures semantic relationships and displays them in an interpretable visual format. Our experimental results demonstrate the method's effectiveness in identifying deep structural similarities between code fragments.

This approach has implications for improving code understanding, ensuring code quality, and improving software development processes. Our method contributes to developing more efficient, maintainable, and secure software systems by facilitating tasks such as automated code review, refactoring, and plagiarism detection. Future research will aim to expand the applicability and scalability of this method, potentially integrating it into broader code analysis and software engineering toolchains.

Additional future directions include the development of more language-agnostic models and improvements in the integration of structural information. Furthermore, exploring the combination of machine learning-based approaches with traditional program analysis techniques could lead to more accurate models for code similarity detection. However, challenges remain in generalizing across different programming languages and coding styles and addressing these will be a crucial focus of ongoing research efforts.

\section*{Acknowledgments}
The research reported in this paper has been funded by the Federal Ministry for Climate Action, Environment, Energy, Mobility, Innovation, and Technology (BMK), the Federal Ministry for Digital and Economic Affairs (BMDW), and the State of Upper Austria in the frame of SCCH, a center in the COMET - Competence Centers for Excellent Technologies Programme.

\bibliography{mybib}

\end{document}